\documentclass[12pt]{article}
\usepackage{cite}
\usepackage{epsfig} 
\usepackage{axodraw}
\usepackage{amssymb}

\def\pr#1{#1^\prime}
\def\ltap{\;\raisebox{-.4ex}{\rlap{$\sim$}} \raisebox{.4ex}{$<$}\;}

\def\beq{\begin{equation}}
\def\eeq{\end{equation}}

\def\beeq{\begin{eqnarray}}
\def\eeeq{\end{eqnarray}}

\newcommand\as{\alpha_{\mathrm{S}}}
\newcommand\asp{\frac{\alpha_{\mathrm{S}}}{\pi}}
\def\ep{\epsilon}
\def\ee{$e^+e^-$}
\def\to{\rightarrow}
\def\nn{\nonumber}
\def\naive{na\"{\i}ve}
\def\qt{q_{\perp}}
\def\q2t{{\bf q}_{\perp}^2}

\def\msbar{\overline{\mbox{\scriptsize MS}}}
\def\ms{$\overline{{\rm MS}}$}

\begin{document}

\begin{titlepage}
\renewcommand{\thefootnote}{\fnsymbol{footnote}}
\begin{flushright}
     CERN--TH/2001-174\\
     UPRF-2001-11\\
     hep-ph/0107138
     \end{flushright}
\par \vspace{10mm}
\begin{center}
{\Large \bf
Soft-Gluon Resummation for the Fragmentation \\ [1ex]
of Light and Heavy Quarks at Large
$x$~\footnote{This work was supported in part 
by the EU Fourth Framework Programme ``Training and Mobility of Researchers'', 
Network ``Quantum Chromodynamics and the Deep Structure of
Elementary Particles'', contract FMRX--CT98--0194 (DG 12 -- MIHT).}}
\end{center}
\par \vspace{2mm}
\begin{center}
{\bf \large Matteo Cacciari}$^{(1)}$ \quad and \quad
{\bf \large Stefano Catani}$^{(2)}$~\footnote{On leave of absence 
from INFN, Sezione di Firenze, Florence, Italy.}\\

\vspace{5mm}
$^{(1)}$~ Dipartimento di Fisica,
Universit\`a di Parma, Italy, and\\
INFN, Sezione di Milano, Gruppo Collegato di Parma\\[10pt]
$^{(2)}$~{CERN - Theory Division\\
CH 1211 Geneva 23, Switzerland} \\

\end{center}

\par \vspace{2mm}
\begin{center} {\large \bf Abstract} \end{center}
\begin{quote}
\pretolerance 10000
\tolerance 10000

We present a QCD study of fragmentation processes for light and heavy
quarks in the
semi-inclusive region of large $x$. Large logarithmic 
terms, due to soft-gluon radiation, 
are evaluated and resummed to all perturbative orders in the QCD coupling 
$\as$. 
Complete analytical results to next-to-leading logarithmic
accuracy are given for one-particle and two-particle inclusive distributions 
in $e^+e^-$ annihilation and 
DIS.
Factorization of parton radiation at low transverse momenta
is exploited to identify the universal (process-independent) perturbative
fragmentation function that controls heavy-quark processes, and to perform
next-to-leading logarithmic resummation of its soft-gluon contributions.
To gauge the  
quantitative impact of resummation, we perform numerical studies of 
light- and heavy-quark fragmentation in
$e^+e^-$ collisions.

\end{quote}

\vspace*{\fill}
\begin{flushleft}
    CERN--TH/2001-174 \\  July 2001

\end{flushleft}
\end{titlepage}

\renewcommand{\thefootnote}{\fnsymbol{footnote}}
\section{Introduction}
\label{sec:intro}

Large logarithmically-enhanced corrections due to soft-gluon radiation
are a general feature in the perturbative-QCD study of hard-scattering 
processes near threshold. Techniques for resumming these corrections
have been developed over the past several years and have been mainly applied to
the production cross sections of high-mass systems.
An extensive and updated list of references can be found in Sect.~5 of
Ref.~\cite{Catani:2000jh}.

In spite of the large amount of available data from \ee, lepton--hadron
and hadron--hadron collisions (see \cite{Webber:2000ui,expdata} and references
therein), soft-gluon effects in 
single-particle (and double-particle) inclusive cross sections have instead
received less attention.
In this paper we consider soft-gluon resummation for the fragmentation 
processes of light- and heavy-flavoured hadrons in the vicinity of the 
threshold region.

The basis for higher-order calculations in perturbative QCD is provided by the
factorization theorem of mass singularities \cite{book}. According to it,
any inclusive hard cross section $\sigma(x,Q^2)$, involving initial-state 
hadrons and detected final-state hadrons, can be written as follows
\beq
\label{fac}
\sigma (x,Q^2) =F\; \otimes \;{\hat \sigma} \; \otimes 
\; D \;+ \;{\cal O}\left( \left( \Lambda/Q \right)^p \right)\;\;\;.
\eeq
The notation in Eq.~(\ref{fac}) is symbolic: $Q^2$ is the hard scale, i.e. a 
typical transferred momentum much larger than the QCD scale $\Lambda^2$; $x$ 
stands for any ratio of other kinematic invariants; the symbol 
$\otimes $ denotes appropriate convolutions of
longitudinal- and transverse-momentum variables and the sum over parton indices
is understood. 
The term ${\cal O}\left( \left( \Lambda/Q \right)^p \right)$ on the right-hand
side of Eq.~(\ref{fac}) stands for cross section contributions that are 
suppressed by some inverse power $p \;(p \geq 1)$ of $Q$ in the 
hard-scattering regime $Q \gg \Lambda$.

Perturbation theory allows us to evaluate the first term on the 
right-hand side of Eq.~(\ref{fac}), that is, the so-called leading-twist 
component of the hard-scattering process. Performing a power series expansion 
in the strong coupling $\as(Q^2)$, we can 
compute the partonic cross section ${\hat \sigma}$ and 
the $Q^2$-evolution of the parton distribution functions $F(x,Q^2)$ and of the
parton fragmentation functions $D(x,Q^2)$.
Note, in particular, that only the $Q^2$-dependence 
(or, more precisely, the anomalous dimensions)
of the parton distributions $F(x,Q^2)$ and $D(x,Q^2)$ is under control 
within perturbation theory. Their absolute normalization at a given (and
arbitrary) scale has to be provided as phenomenological input. In the
perturbative calculation this arbitrariness is reflected by the fact that 
anomalous dimensions and coefficient functions are separately dependent
on the factorization scheme. Any definite prediction thus requires 
a consistent evaluation of anomalous dimensions and coefficient functions 
within the same factorization scheme.

As long as all the kinematic scales are of the same order (i.e. $x =
{\cal O}(1)$), perturbative calculations to the first few orders in the QCD coupling
$\as$ should provide reliable and accurate theoretical predictions
for the hadronic cross section.
However, the perturbative series for the anomalous dimensions {\em and} for the 
coefficient functions are poorly convergent in the semi-inclusive or 
Sudakov region, that is, when the energy or transverse momentum
of the triggered hadron (final state) is a large fraction, $x \to 1$, of the
available energy $\sqrt s$ in the scattering process. In this case the 
production threshold is approached and the emission of accompanying radiation
is strongly inhibited by the kinematics. Only soft particles can be radiated in
the inclusive final state, and the bremsstrahlung spectrum of soft 
(and collinear) gluons produces large 
logarithmic contributions of the type $\as^n \ln^m(1-x)/(1-x)$
(with $m \leq 2n-1$) to each order $n$ in perturbation theory.
In the presence of these contributions, the `true' expansion 
parameter is no longer $\as$ but rather the large effective coupling 
$\as \ln^2(1-x)$ and, hence, {\em any} finite-order perturbative 
calculation is unable to provide an accurate evaluation of the cross section. 
The only reliable procedure consists in resumming classes of logarithms to 
all orders in $\as$. 

Leading and next-to-leading logarithmic contributions to parton distributions
and parton fragmentation functions are known. 
To be definite, let us consider the
\ms\ factorization scheme and introduce the 
$N$ moments\footnote{In $N$-moment space, the semi-inclusive
region $x \to 1$ corresponds to the limit $N \to \infty$.}
$D_{a/h, \,N}(Q^2)$ of the fragmentation function of the parton $a$ into the
light hadron $h$:
\beq
\label{dan}
D^{(\msbar)}_{a/h, \,N}(Q^2) \equiv 
\int_0^1\;dx\;x^{N-1} \;D^{(\msbar)}_{a/h}(x,Q^2) \;\;.
\eeq
The $Q^2$-evolution of the parton fragmentation functions is given by the
Altarelli--Parisi (AP) equations
\beq
\label{gan}
\frac {d D^{(\msbar)}_{a/h, \,N}(Q^2)} {d \ln Q^2} = 
\sum_b \gamma_{ab, \, N}(\as(Q^2)) \; D^{(\msbar)}_{b/h, \,N}(Q^2) \;,
\eeq
where the anomalous dimensions $\gamma_{ab, \, N}$ are the $N$ moments of the
AP probabilities. An important feature of the 
\ms\ factorization scheme is that only the flavour-diagonal\footnote{The 
non-diagonal terms are suppressed by a relative factor of ${\cal O}(1/N)$.}
contributions $\gamma_{qq}$ and $\gamma_{gg}$ to the evolution are 
affected by enhanced logarithmic corrections at large $N$ (or, equivalently,
at large $x$). In particular, the explicit expressions of the 
flavour-diagonal anomalous dimensions are \cite{cfp}
\beeq
\label{gnq}
\gamma_{qq, \,N}(\as) &\simeq& -\;C_F \;\frac{\as}{\pi} \;\left( 1 + K\;
\frac{\as}{2\pi} +{\cal O}(\as^2) \right) \ln N \; + {\cal O}(1) \;\;, \\
\label{gng}
\gamma_{gg, \,N}(\as) &\simeq& -\;C_A \;\frac{\as}{\pi} \;\left( 1 + K\;
\frac{\as}{2\pi} +{\cal O}(\as^2) \right) \ln N \; + {\cal O}(1) \;\;, 
\eeeq
where the coefficient $K$ \cite{kt1} is given by\footnote{In this
paper $\as(Q^2)$ denotes the QCD coupling in the \ms\ renormalization 
scheme. The value of the coefficient $K$ reported in Eq.~(\ref{kcoef}) refers 
to this renormalization scheme. One can also introduce \cite{cmw} an 
alternative renormalization scheme such as to absorb the coefficient $K$ in the 
redefinition of $\as$ according to $\as \to \as (1+K\as/2\pi)$.} 
\beq
\label{kcoef}
K=\;C_A\;\left( \frac{67}{18}- \frac{\pi^2}{6}\right) - \frac{5}{9}\;N_f \;\;,
\eeq
and the term ${\cal O}(1)$ denotes any non-singular contribution at large $N$.
The expressions (\ref{gnq}) and (\ref{gng}) also show another important feature
of the \ms\ factorization scheme. In this scheme the anomalous dimensions
are not more singular than a {\em single} power of $\ln N$ when 
$N \to \infty$ \cite{ste,ct2,gk}.

A similar result is valid for the parton distributions
$F^{(\msbar)}(x,Q^2)$,
and the corresponding anomalous dimensions, in the large-$N$ limit, are again 
given by Eqs.~(\ref{gnq}) and (\ref{gng}).

The knowledge of  the large-$N$ behaviour of the anomalous dimensions is 
however not sufficient to evaluate the
hadronic cross section in the large-$x$ region.
The consistency of the resummation procedure with the factorization theorem of 
mass singularities demands also the calculation, to 
the same logarithmic accuracy, of the process-dependent partonic cross section.
The latter can strongly affect the hadronic cross section because 
(unlike the anomalous dimensions) the perturbative series
for its $N$ moments ${\hat \sigma}_N$ contains {\em double}-logarithmic terms
$\as \ln^2N$ in the general form
\beq
\label{gensig}
{\hat \sigma}_N \sim {\hat \sigma}_N^{(LO)} \left\{ 1 + \sum_{n=1}^{\infty}
\as^n \sum_{m=1}^{2n} c_{n,m} \ln^m N \right\} \;\;,
\eeq
where ${\hat \sigma}_N^{(LO)}$ is the leading-order (LO) contribution.
Moreover, we should keep in mind that the soft-gluon contributions to
the partonic cross section can be sizeable long before the 
threshold region in the hadronic cross section is actually approached.
This is because the evolution of the parton densities and fragmentation
functions sizeably reduces the energy that is available in the 
partonic hard-scattering subprocess; thus, the partonic cross section 
$\hat \sigma$ in the factorization formula
(\ref{fac}) is typically evaluated much closer to threshold than the 
hadronic cross section.

In recent years, the general methods developed in 
Refs.~\cite{ste,ct2,Kidonakis:1997gm,Bonciani:1998vc,Laenen:1998qw,Catani:1998tm} 
have been applied to carry out soft-gluon resummation
to next-to-leading logarithmic accuracy for several processes. Nonetheless,
no explicit resummed calculation has been performed for fragmentation cross
sections.

In this paper, we consider fragmentation cross sections in 
$e^+e^-$ annihilation and deep inelastic lepton--nucleon scattering (DIS).
We resum leading and next-to-leading 
soft-gluon contributions to one-particle and two-particle inclusive
distributions. In the case of the one-particle distribution in \ee, we also
discuss some of the next-to-next-to-leading terms.

We also consider the fragmentation of heavy quarks. In the limit when the
heavy-quark mass $m$ is much smaller than the hard scale $Q$ of the scattering
process, the factorization formula (\ref{fac}) can be generalized in a
process-independent way to compute heavy-quark cross sections. 
The generalization is based on the perturbative fragmentation function 
formalism \cite{Mele:1991cw,Cacciari:1994mq}, which uses the
AP evolution equations (\ref{gan}) to resum the 
large single-logarithmic contributions $(\as \ln Q^2/m^2)^n$ of collinear 
origin. We shall show how
mass and Sudakov effects can systematically be included in the 
AP evolution at low transverse momentum, thus extending the formalism of 
Refs.~\cite{Mele:1991cw,Cacciari:1994mq} to include soft-gluon
resummation at large $x$. Our results demonstrate that in the limit
$m/Q \ll 1$, the soft-gluon contributions are process-independent and
can thus be resummed in the universal perturbative component of 
the heavy-quark fragmentation function. In particular, we generalize
the results of the soft-gluon resummed calculations performed in 
Refs.~\cite{Mele:1991cw,Dokshitzer:1996ev} by evaluating the
heavy-quark fragmentation function to next-to-leading logarithmic accuracy
in the large-$x$ region. 

The outline of the paper is as follows. In Section~\ref{section2} we
consider light-quark fragmentation. In Sect.~\ref{1ee} we provide analytical 
results for
the next-to-leading Sudakov resummation of the \ee\ coefficient function
in the $\overline{\rm MS}$ scheme. In the same section we also discuss
some next-to-next-to-leading terms, and point out a universality pattern
by comparing with the DIS structure function case. In Sect.~\ref{sec:numL} 
we then perform
some numerical studies and assess the impact of the resummation on the
value of the \ee\  
single-inclusive distribution at large $x$, and on the
stability of the result with respect to renormalization/factorization
scale variations.
In Section~\ref{section3} we consider the case of heavy-quark
fragmentation. We start in Sect.~\ref{sec:CR} by briefly reviewing the 
perturbative fragmentation function formalism.
In Sect.~\ref{sec:qcolfac} we show how the process-independent initial 
condition  
for the perturbative fragmentation function can be introduced by exploiting
the universal factorization properties of parton radiation at low transverse
momenta. Next-to-leading Sudakov resummation for the initial condition is
performed in Sect.~\ref{sec:resHQ}. The numerical studies performed in 
Sects.~\ref{sec:resHQ} and \ref{sec:eeHQ}
assess the effects of Sudakov resummation on softening and scale dependence 
of the 
fragmentation function and single-inclusive \ee\
distribution of heavy quarks.
In Sect.~\ref{sec:sum} we finally summarize our main results.
Soft-gluon resummation for two-particle distributions in \ee\
collisions and one-hadron inclusive cross section in DIS is considered 
in Appendices A and B, respectively.

\section{Light-quark fragmentation 
at large $x$}
\label{section2}
\vskip 10pt

In this section we consider the fragmentation of light quarks (light hadrons)
by performing a detailed theoretical and numerical study of the single-particle
inclusive cross section in $e^+e^-$ collisions. Related theoretical results
on the two-particle distribution in \ee\ annihilation and the single-inclusive
cross section in DIS are presented in the appendices.

\subsection{Single-particle inclusive distribution in $e^+e^-$ annihilation}
\label{1ee}

\begin{figure}[t]
\begin{center}
\begin{picture}(130,100)(0,0)         
\ArrowLine           ( 0,10)( 40,50)  
\ArrowLine           ( 0,90)( 40,50)
\Photon              ( 40,50)( 90,50){3}{4}   
\ArrowLine                (90,50)(120,90)
\ArrowLine                (90,50)(120,20)
\Line                (90,50)(115,15)
\ArrowLine                (90,50)(110,10)
\BCirc               ( 90,50){10}
\Text                ( 120, 15)[l]{ $X$}
\Text                ( 125,90)[l]{$h(p)$}
\Text                ( 45,65)[l]{$V(Q)$}
\Text                ( 5,65)[l]{$e^+$}
\Text                ( 5,35)[l]{$e^-$}
\end{picture}
\caption{\label{epemfig}\small Inclusive production of a hadron $h$ with 
momentum $p$ in $e^+e^-$ annihilation.
}
\end{center}
\end{figure}
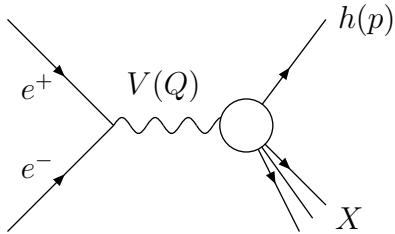

We consider the inclusive production of a single-hadron $h$ with
momentum $p$ in $e^+e^-$ annihilation.
Within the single-vector boson ($V=\gamma^*,Z^0$) exchange
approximation (Fig.~\ref{epemfig}),
\beq
\label{gammapx}
e^+ + e^- \to V(Q)\to h(p) + X \;\;,
\eeq
the single-particle angular distribution
has three (transverse, asymmetric and longitudinal) components 
\cite{aemp,Nason:1994xx,RvN}.
Each component can then be split in flavour singlet and 
flavour non-singlet contributions. In the following we do not consider the
longitudinal cross section or the `pure' flavour-singlet contributions,
because they are suppressed
by a relative factor of the order of $(1-x)$ when $x \to 1$.
The flavour non-singlet contributions can be written as
\beq \label{dshad}
\frac{d\sigma_h^{(e^+e^-)}(x,Q^2)}{dx} = \sigma^{(LO)} 
\;\int_x^1 \frac{dz}{z}   
\;C^{(e^+e^-)\msbar}(x/z,\as(\mu^2);Q^2,\mu^2,\mu_F^2)
\;D^{(\msbar)}(z,\mu_F^2) \;\;,
\eeq
where $x$ is the fraction of the beam energy carried by the hadron,
\beq
\label{x}
x=\frac{2p \cdot Q}{Q^2} \;\;,
\eeq
and $\mu$ and $\mu_F$ denote the renormalization and factorization scales,
respectively.
The expression (\ref{dshad}) is valid for both the transverse and asymmetric
cross sections: $\sigma^{(LO)}$ are the corresponding cross sections at LO
and $D^{(\msbar)}$ are the corresponding flavour non-singlet components of
the quark fragmentation functions into the hadron $h$. In the following we
consider the large-$x$ behaviour of the (flavour non-singlet) 
coefficient function 
$C^{(e^+e^-)\msbar}(x)$, and we do not make any distinction between transverse
and asymmetric coefficient functions because they only differ by terms that
are not singular in the limit $x \to 1$.

The perturbative calculation of the coefficient function\footnote{From now on,
we omit the
label $\msbar$ and we always use the \ms\ factorization scheme,
unless explicitly stated otherwise.}
gives \cite{aemp}:
\beeq
\label{dxqq0}
&&\!\!\!\!\!\!C^{(e^+e^-)}(x,\as(\mu^2);Q^2,\mu^2,\mu_F^2)
=\delta(1-x) \;+\; \frac{\as(\mu^2)}{\pi}C_F\;\left[ \frac{1}{2}
\left(\frac
{1+x^2}{1-x}\right)_+ \ln\frac{Q^2}{\mu_F^2} \right. \nonumber \\
&&\;\;\;\;\;\; + \left( \frac{\ln (1-x)}{1-x}\right)_+  - \left.
\frac{3}{4}\left( \frac{1}{1-x} \right)_+ +\left( \frac{\pi^2}{3}-\frac
{9}{4}\right) \delta(1-x) + f(x)\right] 
+ {\cal O}(\as^2) \;\;,
\eeeq
where $f(x)$ is a smooth function\footnote{To be precise,
$f(x) \sim \ln(1-x)$ when $x \rightarrow 1$, and thus the contribution of
$f(x)$ to the right-hand side of Eq.~(\ref{dxqq0}) is less singular than that
of the other distributions in the square bracket.} for $x \rightarrow 1$ and the  
$(\dots)_+$-distributions are defined in the customary way:
\beq
\label{+}
\int_0^1 dz\;h(z) \;[\, g(z) \,]_+  \equiv \int_0^1 dz \;[h(z)-h(1)]\;g(z)
\;.
\eeq
Introducing the $N$ moments, as in Eq.~(\ref{dan}), the expression in  
Eq.~(\ref{dxqq0}) reads
\beeq
\label{dnqq0}
&&C_N^{(e^+e^-)}(\as(\mu^2);Q^2,\mu^2,\mu_F^2)
= 1 + \frac{\as(\mu^2)}{\pi}C_F\; \left[\;-\left( \ln N+\gamma_E- 
\frac{3}{4} \right)
\;\ln\frac{Q^2}{\mu_F^2} + \frac{1}{2} \ln^2N  \right. \nonumber  \\
&&\;\;\;\;\;\; + \left. \left (\frac{3}{4}+\gamma_E \right) \ln N +
\left( \frac{5}{12}\pi^2-\frac{9}{4}+
\frac{1}{2}\gamma_E^2+\frac{3}{4}\gamma_E \right) + {\cal O}\left(
\frac{1}{N} \right) \right] + {\cal O}(\as^2) \;\;,
\eeeq
where $\gamma_E = 0.5772\dots$ is the Euler constant.
The large (when $N \to \infty$) contributions $\ln^{k+1}N$ in 
Eq.~(\ref{dnqq0}) are due to the Mellin transformation of the 
singular (when $x \to 1$) distributions $[\ln^k(1-x)/(1-x)]_+$, associated
to the bremsstrahlung spectrum of soft and collinear emission.
To higher orders in perturbation theory, the large-$N$ moments of the 
coefficient function have the general double-logarithmic expansion on the
right-hand side of Eq.~(\ref{gensig}).

To perform the all-order resummation of the large $\ln N$ contributions,
we can apply the general method developed in Ref.~\cite{ct2}.
The single-particle distribution in 
Eq.~(\ref{dshad}) is obtained by crossing to the final state 
the momentum of the incoming quark in the DIS process. 
Thus, we repeat step by step and in a straightforward manner 
the calculation of the DIS cross section carried out in Ref.~\cite{ct2};
we obtain the following resummed expression for the \ee coefficient
function:
\beeq
\ln C_N^{(e^+e^-)}(\as(\mu^2);Q^2,\mu^2,\mu_F^2) &=&
\ln \Delta_N(\as(\mu^2),Q^2/\mu^2;Q^2/\mu_F^2) \nn \\
\label{cnee}
&+& \ln J_N(\as(\mu^2),Q^2/\mu^2) + {\cal O}(\as(\as \ln N)^k) \;\;.
\eeeq
The radiative factors $\Delta_N$ and $J_N$ have the following
{\em exponentiated} form:
\beq 
\Delta_N(\as(\mu^2),Q^2/\mu^2;Q^2/\mu_F^2) = \exp \left\{
\int_0^1 dz \;\frac{z^{N-1} -1}{1-z} \;
\int_{\mu_F^2}^{(1-z)^2Q^2} 
\frac{dq^2}{q^2} \;A[\as(q^2)] \right\} \,,
\label{deltams}
\eeq
\beeq
J_N(\as(\mu^2),Q^2/\mu^2) &=& \exp \Bigg\{
\int_0^1 dz \;\frac{z^{N-1} -1}{1-z} 
\;\Big[ \int_{(1-z)^2Q^2}^{(1-z)Q^2} 
\frac{dq^2}{q^2} \;A[\as(q^2)]
 \nonumber \\ &&
 + \, \frac{1}{2} \;B[\as((1-z)Q^2)] \Big] \Bigg\} \;,
\label{jfun}
\eeeq
where the functions $A(\as)$ and $B(\as)$ have perturbative expansions in
$\as$,
\beeq
\label{afun}
A(\as)&=& \sum_{n=1}^{\infty} \left( \asp \right)^n \;A^{(n)} \;\;, \\
\label{bfun}
B(\as)&=&\sum_{n=1}^{\infty} \left( \asp \right)^n \;B^{(n)} \;\;,
\eeeq
whose first coefficients are:
\beeq
\label{a1a2}
A^{(1)}&=&C_F\;\;, \;\;\; A^{(2)}= \frac{1}{2} \,C_F\;
K = \frac{1}{2}\;C_F\;\left[ C_A \left(\frac{67}{18}- \frac{\pi^2}{6}\right)
- \frac{5}{9}N_f \right] \;\;, \\
\label{b1} 
B^{(1)}&=& -\frac{3}{2}\, C_F \;\;. 
\eeeq
Performing the integrations over $z$ and $q^2$ in 
Eqs.~(\ref{deltams}) and (\ref{jfun}), we obtain a series 
of logarithmic contributions of the type $\as^n \ln^m N$, with $m \leq (n+1)$.
We define as {\em leading} logarithmic (LL) the terms with $m = n+1$.
The {\em next-to-leading} logarithmic (NLL) contributions are those with 
$m=n$, the {\em next-to-next-to-leading} logarithmic (NNLL) terms have $m=n-1$,
and so forth.

Owing to the knowledge of the LL coefficient $A^{(1)}$ and of the NLL
coefficients $A^{(2)}, B^{(1)}$ in Eqs.~(\ref{a1a2}) and  (\ref{b1}),
the result in 
Eq.~(\ref{cnee}) resums all the leading and next-to-leading 
$\ln N$ contributions to the $N$ moments of the coefficient function
of the single-particle distribution in \ee\ annihilation.

Note that, according to 
Eqs.~(\ref{cnee}) and (\ref{deltams}), 
in the large-$N$ limit, the \ee\ coefficient function
has the following dependence on the factorization scale $\mu_F$:
\beq
\label{mufdep}
\frac{d \ln C_N^{(e^+e^-)}(\as(\mu^2);Q^2,\mu^2,\mu_F^2)}{d \ln \mu_F^2}
\simeq - A[\as(\mu_F^2)] \int_0^1 dz \;\frac{z^{N-1} - 1}{1-z} \;\;.
\eeq
Since, to the logarithmic accuracy of Eqs.~(\ref{cnee}) and (\ref{gnq}),
the (non-singlet) quark anomalous dimension $\gamma_{qq, N}(\as)$
can be written as
\beq
\label{andim}
\gamma_N(\as) \simeq A(\as) \int_0^1 dz \;\frac{z^{N-1} - 1}{1-z} \simeq
 - \;A(\as) \;[ \; \ln N + {\cal O}(1) \;] \;\;,
\eeq
Eq.~(\ref{mufdep}) shows that the $\mu_F$-dependence of the coefficient 
function 
$C_N^{(e^+e^-)}$ consistently matches (and thus cancels)
the $\mu_F$-dependence of the fragmentation function $D_N^{(\msbar)}$ in
Eq.~(\ref{dshad}).

The NLL resummed result in Eq.~(\ref{cnee}) has a simple physical
interpretation \cite{cmw,ct2} in terms of independent fragmentation of 
the observed hadron $h$ and of the recoiling jet.
The radiative factor $\Delta_N$ describes the energy loss of the primary
quark (or antiquark) that eventually fragments into the triggered hadron.
This factor takes into account final-state radiation of gluons that are soft
(i.e. with energy fraction $\omega/p_0 \equiv 1-z \leq 1 - x \sim 1/N \ll 1$)
{\em and} collinear (i.e. with small transverse momentum 
$q \sim \omega \theta=(1-z)p_0 \theta \ll (1-z)Q$) with respect
to the momentum $p^\mu$ of the observed hadron. Having fixed the energy
of the jet that produces the observed hadron, the recoiling jet is constrained
to carry a small invariant mass squared $k^2=(1-z)Q^2 \ltap (1 - x)Q^2 \ll 1$.
The radiative factor $J_N$ describes the fragmentation of the invariant mass
of the recoiling jet as produced by the final-state radiation of collinear
(either soft or hard) partons.

This independent-fragmentation picture is an effective physical picture.
Although the two jets do not evolve independently (classically), the 
quantum interferences due to non-collinear parton radiation cancel
up to NLL accuracy. The destructive interference of soft-parton radiation
at large angles follows from QCD coherence \cite{coher}, 
but it is no longer complete beyond NLL order. The resummed
expression in Eq.~(\ref{cnee}) can be extended to any logarithmic order
as \cite{Kidonakis:1997gm,Bonciani:1998vc,Laenen:1998qw,Catani:1998tm}
\beeq
\label{eeall}
C_N^{(e^+e^-)}(\as(\mu^2);Q^2,\mu^2,\mu_F^2) &=&
c(\as(\mu^2),Q^2/\mu^2;Q^2/\mu_F^2) \;
\Delta_N^{({\rm int})}(\as(\mu^2),Q^2/\mu^2) \\
&\cdot& \Delta_N(\as(\mu^2),Q^2/\mu^2;Q^2/\mu_F^2) 
\; J_N(\as(\mu^2),Q^2/\mu^2) + {\cal O}(1/N) \;\;. \nn 
\eeeq
This expression contains two other factors in addition to those
of Eq.~(\ref{cnee}). 
The factor $c(\as)$ does not depend on $N$, and it is due
to radiative corrections produced by hard (with energy $E \sim Q$)
virtual partons. This factor is computable as a power series expansion in
$\as$
\beq
c(\as(\mu^2),Q^2/\mu^2;Q^2/\mu_F^2) = 1 + \sum_{n=1}^{\infty}
\left( \asp \right)^n \;c^{(n)}(Q^2/\mu^2;Q^2/\mu_F^2) \;\;.
\eeq
The radiative factor $\Delta_N^{({\rm int})}$ is given by
\beq
\label{dint}
\Delta_N^{({\rm int})}(\as(\mu^2),Q^2/\mu^2) = \exp \Bigg\{
\int_0^1 dz \;\frac{z^{N-1} -1}{1-z} 
\;D[\as((1-z)^2Q^2)]  \Bigg\} \;,
\eeq
where
\beq
\label{dfun}
D(\as) = \left( \asp \right)^2 \;D^{(2)} + 
\sum_{n=3}^{\infty} \left( \asp \right)^n \;D^{(n)} \;\;.
\eeq
Note that $\Delta_N^{({\rm int})}$ embodies $\ln N$ contributions, and that
the perturbative function $D(\as)$ in Eq.~(\ref{dfun}) has a vanishing
first-order coefficient $D^{(1)}$. Thus, $\Delta_N^{({\rm int})}$ contributes
to the \ee\ coefficient function $C_N^{(e^+e^-)}$ starting from
NNLL order. This radiative factor takes into account soft-parton
radiation at large angle (or, equivalently, with transverse momenta 
$q\sim \omega \theta \sim (1-z) Q$) and
it leads to violation  
(at NNLL accuracy) of the `independent-fragmentation' picture discussed above.

It is straightforward to check that our NLL result in Eq.~(\ref{cnee})
agrees with the large-$N$ limit of the exact ${\cal O}(\as^2)$ calculation
of Ref.~\cite{RvN}. Moreover, this calculation can also be used to extract 
a linear combination of the NNLL coefficients $B^{(2)}$ and $D^{(2)}$ in 
Eqs.~(\ref{bfun}) and (\ref{dfun}). We find
\beeq
\label{nnllc}
D^{(2)} + \frac{1}{2} \,B^{(2)} &=& \frac{1}{16} \left[
C_F^2 \left( - \frac{3}{2} + 2 \pi^2 - 24 \zeta_3  \right)
+ C_F C_A \left( - \frac{3155}{54} + \frac{22}{9} \pi^2 + 40 \zeta_3 \right)
\right. \nn \\
&+& \left. C_F N_f \left( \frac{247}{27} - \frac{4}{9} \pi^2   \right) \right]
\;\;,
\eeeq
where $\zeta_n$ is the Riemann zeta function ($\zeta_3=1.202\dots$).
An independent calculation of $B^{(2)}$ and $D^{(2)}$ could be performed
by exploiting infrared-factorization formulae 
\cite{infrafac2,infrafac1} at ${\cal O}(\as^2)$ (see, e.g., the
analogous calculations carried out in 
Refs.~\cite{deFlorian:2000pr, Catani:2001ic}).
The knowledge of the remaining NNLL coefficient $A^{(3)}$ in Eq.~(\ref{afun})
requires the ${\cal O}(\as^3)$-calculation of the anomalous dimensions
$\gamma_{qq,N}(\as)$ that control the evolution of the 
\ms\ fragmentation functions.

It is interesting to compare the results in Eqs.~(\ref{cnee}) and (\ref{eeall})
with the corresponding resummed expression,
given in Refs.~\cite{cmw,ct2}, of the coefficient function for the
DIS structure functions at large values of the Bjorken variable $x$. The
comparison shows that, not only the anomalous dimensions of 
the parton distributions and fragmentation functions 
but also the coefficient functions exactly coincide in the semi-inclusive 
limit to NLL accuracy. Moreover, the lowest-order NNLL coefficient
in Eq.~(\ref{nnllc}) also coincides with the corresponding 
coefficient\footnote{Note that the normalization of
our coefficients $B^{(2)}$ and $D^{(2)}$ is different from that of the DIS
coefficients in Ref.~\cite{Vogt:2001ci}. More precisely,
the combination $(D^{(2)} + B^{(2)})_{DIS}$ in Eq.~(22) of 
Ref.~\cite{Vogt:2001ci} corresponds to our combination 
$16(D^{(2)} + B^{(2)}/2)$.}
for the DIS process \cite{Vogt:2001ci}.
This correspondence extends, beyond the leading collinear level,
the validity of the Gribov--Lipatov perturbative relation \cite{gl} 
between DIS structure functions and \ee fragmentation functions.

\setcounter{footnote}{0}
\subsection{Numerical results}
\label{sec:numL}

We present some numerical results to illustrate the main quantitative
effects of soft-gluon resummation on the single-particle distribution
in \ee annihilation. Since the ${\cal O}(\as^3)$ anomalous dimensions
(and the NNLL coefficient $A^{(3)}$) are not known, we limit
ourselves to considering NLL resummation at large-$x$ matched to the complete 
next-to-leading order (NLO) calculation \cite{aemp,Nason:1994xx} 
of the coefficient function.

Introducing the first two coefficients, $b_0$ and $b_1$,
of the QCD $\beta$-function,
\beq
\label{betass}
b_0 = \frac{11 C_A - 4 T_R N_f}{12\pi}\;,\;\;\;\;\; b_1 =
\frac{17 C_A^2 - 10 C_A T_R N_f -6 C_F T_R N_f}{24\pi^2}\;,
\eeq
in terms of which we have\footnote{In our numerical calculations we always
use the
value $\Lambda^{(5)}=200$~MeV for the QCD scale $\Lambda^{(5)}$ with
$N_f=5$ effective massless flavours. This corresponds to $\as(M_Z^2)=0.116$.}
\beq
\as(\mu^2) = {1\over{b_0\ln(\mu^2/\Lambda^2)}} 
\left(1-{{b_1\ln\ln(\mu^2/\Lambda^2)}\over
{b_0^2\ln(\mu^2/\Lambda^2)}}\right)\,,
\eeq
and defining the variable $\lambda$,
\beq
\label{lamdef}
\lambda = b_0 \;\as(\mu^2) \,\ln N \;\;,
\eeq
we first evaluate the radiative factors in Eqs.~(\ref{deltams}) and 
(\ref{jfun}) at NLL accuracy\footnote{This is achieved \cite{ct2} by replacing
$z^{N-1} -1 \to -\Theta\left(1-\frac{e^{-\gamma_E}}{N} - z\right)$ 
in the integrand.} and we obtain:
\beeq
\ln \Delta_N(\as(\mu^2),Q^2/\mu^2;Q^2/\mu_F^2) 
&=& \ln N \;h^{(1)}(\lambda)
\nonumber \\  &+&
h^{(2)}(\lambda,Q^2/\mu^2;Q^2/\mu_F^2) + 
{\cal O}\left(\as(\as \ln N)^k\right) \,,\label{lndeltams} \phantom{aaa} \\
\ln J_N(\as(\mu^2),Q^2/\mu^2) &=& \ln N \;f^{(1)}(\lambda)
\nonumber \\ &+&
f^{(2)}(\lambda,Q^2/\mu^2) + {\cal O}\left(\as(\as \ln N)^k\right) \,.
\label{lnjfun}
\eeeq
The LL and NLL functions $h^{(1)},f^{(1)}$ and $h^{(2)},f^{(2)}$ 
are given, in terms of the perturbative coefficients
$A^{(1)}, A^{(2)}, B^{(1)}, b_0, b_1$, in Eqs.~(75)--(78) of 
Ref.~\cite{Catani:1998tm}.

The Sudakov-resummed part $C_N^{S}$ of the \ee coefficient
function is then written (from Eq.~(\ref{eeall})) at NLL accuracy as
\beeq
&&C_N^{S}(\as(\mu^2);Q^2,\mu^2,\mu_F^2) = 
\left\{ 1 + \frac{\as(\mu^2)}{\pi}C_F\; \left[
\frac{5}{12}\pi^2-\frac{9}{4}+
\frac{1}{2}\gamma_E^2+\frac{3}{4}\gamma_E  
\right. \right. \nn \\ 
&& \quad + \left. \left. 
\left( \frac{3}{4} - \gamma_E  \right) \;\ln\frac{Q^2}{\mu_F^2} 
\right] \right\} \cdot
\exp \Bigl[ \ln N \;g^{(1)}(\lambda) + g^{(2)}(\lambda,Q^2/\mu^2;Q^2/\mu_F^2) 
\Bigr] \;,
\label{cneeres}
\eeeq
where
\beeq
\label{g1funms} 
g^{(1)}(\lambda) &=& h^{(1)}(\lambda) + f^{(1)}(\lambda) =
\frac{A^{(1)}}{\pi b_0 \lambda} \;
\bigl[ \lambda + (1-\lambda) \ln (1-\lambda) \bigr] \;,\\
g^{(2)}(\lambda,Q^2/\mu^2;Q^2/\mu_F^2) 
&=& h^{(2)}(\lambda,Q^2/\mu^2;Q^2/\mu_F^2) +
f^{(2)}(\lambda,Q^2/\mu^2) \nn \\
&=& \frac{A^{(1)}  b_1}{2 \pi b_0^3}
\left[ 2\lambda + 2 \ln (1-\lambda) +  \ln^2 (1-\lambda) \right]
+ \frac{\left( B^{(1)} -2 A^{(1)}\gamma_E \right)}{2 \pi b_0} \ln (1-\lambda)
\nn \\
\label{g2funms}
&-&\frac{1}{\pi b_0} \left[\lambda + \ln (1-\lambda) \right] 
\left( \frac{A^{(2)}}{\pi b_0} - A^{(1)} \ln \frac{Q^2}{\mu^2} \right) 
- \frac{A^{(1)}}{\pi b_0} \;\lambda \;\ln \frac{Q^2}{\mu_F^2} \;. 
\eeeq
and the term in the curly bracket is the constant (when $N \to \infty$) part
of the coefficient function at ${\cal O}(\as)$ (see Eq.~(\ref{dnqq0})).
 
Our final expression for the $N$ moments of the coefficient function 
is 
\beeq
\label{resfin}
C_N^{(\rm res)}(\as(\mu^2);Q^2,\mu^2,\mu_F^2)
&=& C_N^{S}(\as(\mu^2);Q^2,\mu^2,\mu_F^2)
- \left[ C_N^{S}(\as(\mu^2);Q^2,\mu^2,\mu_F^2) \right]_{\as} \nn \\
&+& \left[ C_N^{(e^+e^-)}(\as(\mu^2);Q^2,\mu^2,\mu_F^2) \right]_{\as} \;\;,
\eeeq
where $[ C_N^{(e^+e^-)} ]_{\as}$
is the full \ee coefficient function
at ${\cal O}(\as)$ \cite{aemp,Nason:1994xx}, 
$C_N^{S}$ is given in Eq.~(\ref{cneeres}) and 
$\left[ C_N^{S}\right]_{\as}$ represents its perturbative truncation
at ${\cal O}(\as)$ (i.e. at NLO). Owing to the subtraction applied to the
resummed part $C_N^{S}$ on the right-hand side, Eq.~(\ref{resfin})
exactly reproduces the NLO result and resums soft-gluon effects beyond 
${\cal O}(\as)$ to NLL accuracy. This defines our NLO+NLL resummed calculation.

To obtain the single-particle inclusive cross section in Eq.~(\ref{dshad}),
the $N$ moments of the coefficient function have to be
multiplied by the $N$ moments $D_N(\mu_F^2)$
of the parton fragmentation functions, and then we have to perfom the inverse
Mellin transformation to the 
$x$ space. Note that the resummed
part of the $N$ moments of the coefficient function has cut singularities
that start at the branch-point $N=N_L=\exp(1/b_0 \as)$
(i.e. at $\lambda=1$ in Eqs.~(\ref{g1funms}) and (\ref{g2funms})) 
in the complex variable $N$. These singularities, which are related to the 
divergent behaviour of the running coupling $\as(q^2)$ near the Landau pole at 
$q=\Lambda$, signal the onset of non-perturbative phenomena at very large 
values of $N$ or, equivalently, when $x$ is very close to its 
threshold value $x=1$. As discussed in detail in Ref.~\cite{Catani:1996yz},
if we are not interested in very high (small) values of $x$ $(Q^2)$, 
we can avoid the explicit introduction of non-perturbative effects to deal with
the Landau singularity. We thus use the Minimal Prescription of 
Ref.~\cite{Catani:1996yz}: 
the inverse Mellin transformation is performed in the complex $N$ plane
by choosing an integration contour that has all the singularities on its left,
except for the singularity at $N=N_L$, which should lie far on its right.
We numerically carry out the inverse Mellin transformation along this contour.

\begin{figure}[t]
\begin{center}
\epsfig{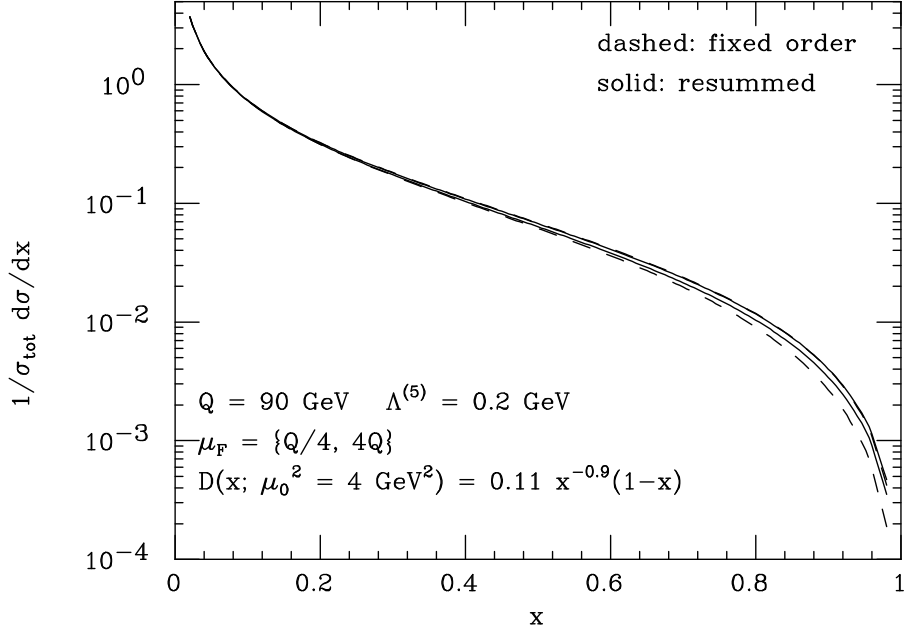}\hfill
\caption{\label{fig:light}\small Effect of Sudakov resummation on a typical 
light-hadron fragmentation spectrum in $e^+e^-$ collisions.
The single-particle distribution is computed at NLO (dashed lines) and
at NLO+NLL accuracy (solid lines). The factorization and
renormalization scales are varied in the range $Q/4 \leq \mu_F=\mu \leq 4Q$. 
}
\end{center}
\end{figure}

\begin{figure}[t]
\begin{center}
\epsfig{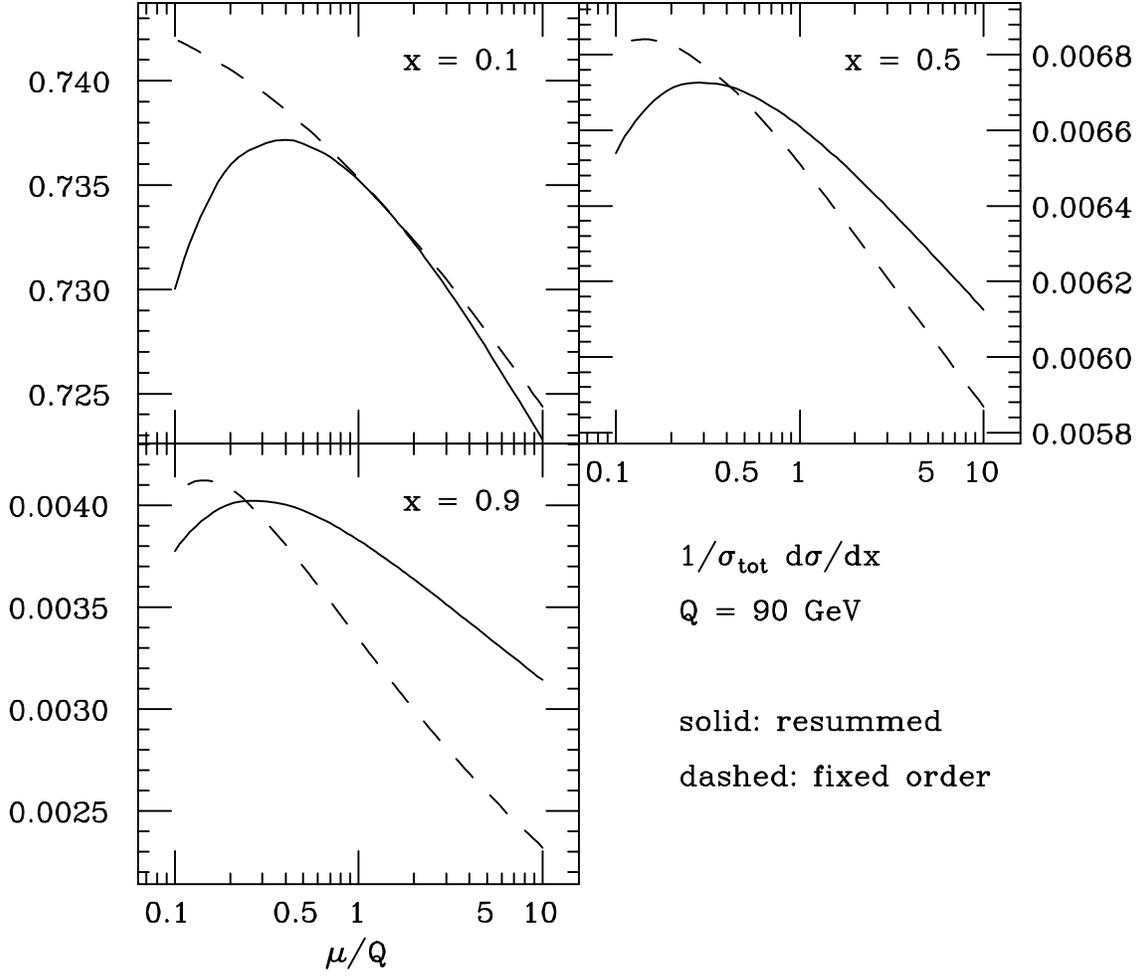}\hfill
\caption{\label{fig:light2}\small
Dependence of the 
light-hadron fragmentation spectrum shown in
Fig.~\protect\ref{fig:light} on the factorization/renormalization scales
$\mu=\mu_F$, at three different valules of $x$.
}
\end{center}
\end{figure}

\begin{figure}[t]
\begin{center}
\epsfig{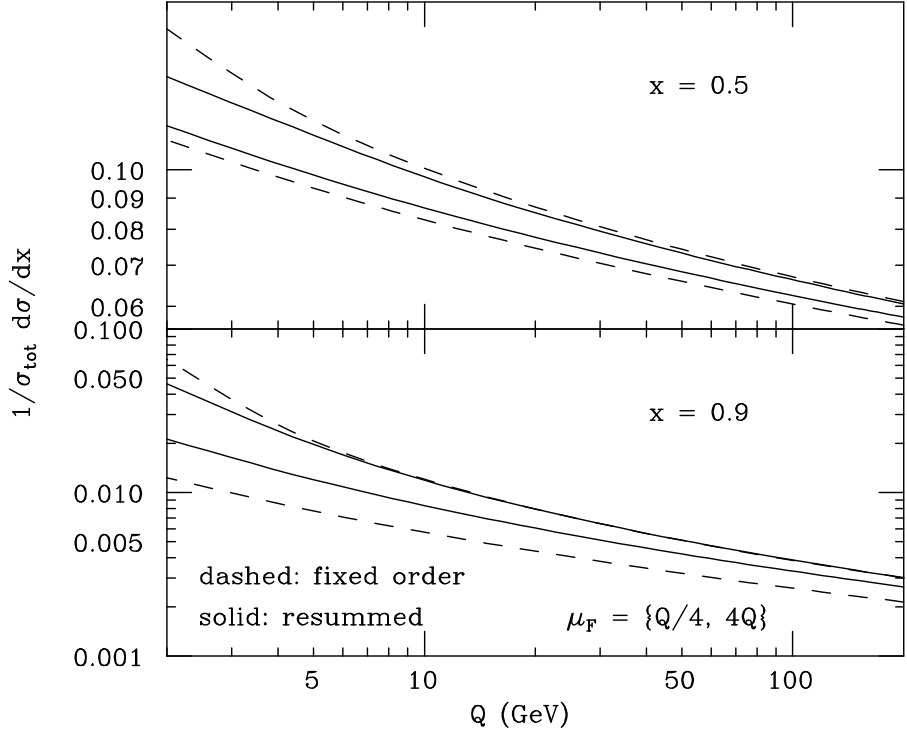}\hfill
\caption{\label{fig:light3}\small
Scaling violations at $x=0.5$ and $x=0.9$, and uncertainty bands due to
variations of the renormalization/factorization scales ($\mu = \mu_F$). 
The fragmentation function at the input scale $\mu_0$ is the same as in
Fig.~\protect\ref{fig:light}.
}
\end{center}
\end{figure}

We present numerical results for the 
single inclusive (transverse plus longitudinal) distribution 
$1/\sigma_{\rm tot} \;d\sigma/dx$ in \ee\ annihilation, where the total 
hadronic cross section $\sigma_{\rm tot}$ is evaluated at NLO:
\beq
\label{stot}
\sigma_{\rm tot}(Q^2) = \sigma^{(LO)}(Q^2) \left[ 1 +\frac{\as(Q^2)}{\pi}
+ {\cal O}(\as^2) \right] \;.
\eeq
For our illustrative purpose, we have considered only the non-singlet
component of the distribution, and we have chosen 
a typical $x$-spectrum \cite{pff},
$D(x;\mu_0^2)=0.11 x^{-0.9}(1-x)$ 
(normalized by $\int_0^1 dx \;D(x;\mu_0^2)=1$), for the fragmentation function
at the input scale $\mu_0 = 2$~GeV. The input fragmentation function is
evolved up to the scale $\mu_F$ by using the NLO AP equation in the
non-singlet channel (see Eq.~(\ref{nloeo}) for details), 
and then it is convoluted with the $e^+e^-$
coefficient function. The (non-singlet) coefficient function is evaluated
either at NLO or at NLO+NLL accuracy. 
The renormalization ($\mu$) and factorization ($\mu_F$) scales are fixed
to be equal and are varied within the range $Q/4 \leq \mu \leq 4Q$.

The NLO and resummed calculations at the centre-of-mass energy $Q=90$~GeV
are compared in Fig.~\ref{fig:light}. As expected, the results of the
two calculations only differ at large $x$. In the large-$x$ region
the resummed calculation has a reduced dependence on the scale $\mu$. 
Moreover, at fixed scale $\mu=Q$ and fixed input of the fragmentation function, 
soft-gluon resummation increases the value of the single-particle
distribution. This resummation effect can be mimicked
by lowering the value of $\mu$ in the NLO calculation.

The enhancement of the distribution in the resummed calculation
can appear surprising,
since the suppression of radiation near threshold is physically expected to
decrease the cross section. The apparent contradiction is due to the actual
definition of the fragmentation function and coefficient function. The effect
of inhibiting soft-gluon radiation is included in part in the AP evolution
of the fragmentation function and in part in the coefficient function, and the
separation depends on the factorization scheme. It turns out that, in the 
\ms\ factorization scheme, the AP evolution overestimates the physical
effect of soft-gluon suppression. This overestimate is included in both the NLO
and NLO+NLL calculations. The additional contributions that are
resummed in the NLO+NLL case are those soft-gluon effects that are left in the
coefficient function after factorization of the evolved fragmentation function.
Having included too much soft-gluon suppression in the \ms\ evolution,
the residual effect in the coefficient function is positive and 
tends to enhance the fixed-order perturbative distribution.

More numerical studies on the effect of resummation are shown in 
Figs.~\ref{fig:light2} and \ref{fig:light3}. 

The plots in Fig.~\ref{fig:light2} show 
the dependence of the fixed-order
(dashed line) and resummed results (solid line) on the
renormalization/factorization scales $\mu=\mu_F$, at the three points
$x=0.1,\;0.5,\;0.9$. As expected from Fig.~\ref{fig:light}, 
the NLO+NLL calculation is more
stable and higher (when $\mu \sim Q$) than the 
NLO one at large values of $x$. It
is worth noting that the agreement of the two calculations at $x=0.1$ 
actually looks
restricted to the region $\mu \simeq Q$. However this statement, and in
general the $\mu$-dependence at $x \ltap 0.1$, can strongly
depend on the detailed shape of the input distribution
$D(x;\mu_0^2)$.

The effect of resummation as a function of the centre-of-mass energy $Q$
is shown in Fig.~\ref{fig:light3}, where uncertainty bands are presented
for the scaling violations of the $e^+e^-$ fragmentation spectrum at $x=0.5$
and $x=0.9$. Lowering the energy $Q$, the difference between the resummed and
fixed-order calculations is enhanced by the increase of $\as(Q)$. The 
bands of the resummed calculation are uniformly narrower, indicating that the
NLO+NLL result is indeed more reliable than the NLO
one.

These results on soft-gluon resummation for the \ms\ coefficient
function in \ee\ collisions can be implemented in global fits
\cite{pff} of \ee\ data to extract parton fragmentation functions.
In particular, the reduced scale dependence of the NLO+NLL calculation can
improve the determination of the quark fragmentation functions at large
$x$.

%
%

\section{Heavy-quark fragmentation}
\label{section3}

In this section we consider the fragmentation of heavy quarks, i.e.
quarks whose mass $m$ is large enough ($m \gg \Lambda$) to  be
considered in the perturbative domain.  The {\sl top} quark is
certainly heavy. Also  the {\sl charm} and {\sl bottom} quarks are
usually considered to satisfy this requirement.

Heavy-quark fragmentation processes can be described by the 
perturbative fragmentation formalism, which
resums collinearly-enhanced perturbative contributions. 
In the large-$x$ region, however, there are also 
logarithmically-enhanced contributions due to soft radiation.
Proper resummation of both classes of logarithms is essential to ensure
the reliability of the theoretical predictions. In the following we first
review the perturbative fragmentation formalism, and then discuss 
soft-gluon resummation.

\subsection{Collinear resummation}
\label{sec:CR}

The fragmentation of heavy quarks is a collinear-safe process, because
perturbative collinear singularities are regularized by the finite value
of the heavy-quark mass $m$. Thus heavy-quark fragmentation cross sections
can unambiguously be computed order by order in QCD perturbation theory.
Nonetheless, when the hard scale $Q$ of the fragmentation process is much 
larger than $m$, the perturbative series contains large 
logarithmic contributions, $\as^m(\as \ln Q^2/m^2)^n$, of collinear 
origin. These logarithmic contributions have to be resummed to higher
perturbative orders, and their resummation can be performed in a
process-independent way  by using
the perturbative fragmentation function 
formalism \cite{Mele:1991cw, Cacciari:1994mq}. The formalism amounts
to generalizing the factorization formula in Eq.~(\ref{fac}). The cross section
$\sigma_{\cal Q}$ for the inclusive production of the heavy quark ${\cal Q}$ 
can be written as
\beq
\label{sighq}
\sigma_{\cal Q}(x,Q;m) = \sum_{a} \int_x^1 \frac{dz}{z} 
\;{\hat \sigma}_a(x/z,\as(\mu^2);Q^2,\mu^2,\mu_F^2) 
\;D_{a/{\cal Q}}(z,\mu_F^2,m^2) \;+{\cal O}((m/Q)^p) \;\;,
\eeq
where ${\hat \sigma}_a$ ($a=q,{\bar q},g$)
are the corresponding partonic cross sections in massless QCD and
$D_{a/{\cal Q}}$ is the pertubative fragmentation function of the (massless) 
parton $a$ into the heavy quark ${\cal Q}$.
The term ${\cal O}((m/Q)^p)$ on the right-hand side 
stands for contributions that are suppressed by some power $p$ $(p \geq
1)$ of $m$ in the kinematic regime $m \ll Q$. These contributions do not 
require all-order resummation and can be computed at a given fixed order
in $\as$. Non-perturbative corrections
of the type $\Lambda/Q$ and $\Lambda/m$ are understood on the right-hand side 
of Eq.~(\ref{sighq}).

Note that $\sigma_{\cal Q}$ denotes a generic fragmentation
cross section produced by lepton \cite{Mele:1991cw,Dokshitzer:1996ev},
hadron \cite{Cacciari:1994mq,Cacciari:1998it}
or photon \cite{Cacciari:1996ej} collisions. 
Thus, ${\hat \sigma}_a$
can implicitly contain convolutions with parton distributions and
fragmentation functions of light hadrons. The variable $x$ in Eq.~(\ref{sighq})
generically denotes the momentum fraction of the heavy quark ($x$ can be either
the energy fraction in \ee annihilation or the transverse-momentum fraction
in hadron and photon collisions).

The resummation of the collinear logarithms of the ratio $Q^2/m^2$ is achieved
by writing the $N$ moments of the perturbative fragmentation function as
\beq
\label{pertev}
D_{a/{\cal Q},N}(\mu_F^2,m^2) = \sum_b 
E_{ab,N}(\mu_F^2,\mu_{0F}^2) 
\;D_{b/{\cal Q},N}^{{\rm ini}}(\as(\mu_0^2);\mu_0^2,\mu_{0F}^2,m^2) \;\;,
\eeq
where $E_{ab,N}(\mu_F^2,\mu_{0F}^2)$ is the evolution operator obtained
by solving the AP equations (\ref{gan}),
\beq
\label{apeo}
\frac {d E_{ab,\,N}(\mu_F^2,\mu_{0F}^2)} {d \ln \mu_F^2} = 
\sum_b \gamma_{ab, \, N}(\as(\mu_F^2)) \;E_{ab,\,N}(\mu_F^2,\mu_{0F}^2) \;\;,
\eeq
with the initial condition 
$E_{ab,\,N}(\mu_{0F}^2,\mu_{0F}^2)= \delta_{ab}$. 
The starting point of the 
perturbative evolution in Eq.~(\ref{pertev}) is set by the scale $\mu_{0F}$, 
which has to be chosen of the same order as $m$,
and by the initial condition $D_{b/{\cal Q},\,N}^{{\rm ini}}$,
which is perturbatively computable as power series in $\as$:
\beq
\label{dinfo}
D_{a/{\cal Q},\,N}^{{\rm ini}}(\as(\mu_0^2);\mu_0^2,\mu_{0F}^2,m^2)
= \delta_{aq} +\sum_{n=1}^{\infty} \as^n(\mu_{0}^2) 
\;D_{a/{\cal Q},\,N}^{{\rm ini}\,(n)}(\mu_0^2,\mu_{0F}^2,m^2) \;\;.
\eeq
Note that, by analogy with the partonic cross sections ${\hat \sigma}_a$
in Eq.~(\ref{sighq}), the perturbative expansion of the initial condition
$D_{a/{\cal Q},\,N}^{{\rm ini}}$ depends on the factorization scale $\mu_{0F}$
and on the renormalization scale $\mu_0$. 

The process-dependence of the cross section $\sigma_{\cal Q}(x,Q;m)$
is entirely embodied in the massless partonic cross 
sections ${\hat \sigma}_a$. The perturbative fragmentation function 
$D_{a/{\cal Q},N}(\mu_F^2,m^2)$ (as well as $E_{ab,N}$ and 
$D_{a/{\cal Q},\,N}^{{\rm ini}}$) is instead universal (process-independent)
and can be evaluated once for all.
The resummation of the large collinear logarithms is 
obtained by perturbatively solving the AP equations in Eq.~(\ref{apeo})
and by setting $\mu_F \sim \mu \sim Q$ and  $\mu_{0F} \sim \mu_0 \sim m$.
For instance, we can consider the second-order expansion of the
flavour non-singlet component $\gamma_{qq,N}$
of the anomalous dimensions,
\beq
\gamma_{qq,N}(\as) = {{\as}\over{2\pi}}
\left[ P_N^{(0)} + {{\as}\over{2\pi}} P_N^{(1)} + {\cal O}(\as^2)  \right]\, ,
\eeq
where $P_N^{(0)}$ and $P_N^{(1)}$ are the corresponding AP probabilities  
at LO and NLO. The evolution operator $E_{N}(\mu_F^2,\mu_{0F}^2)$ for the
non-singlet channel thus reads
\beeq
\label{nloeo}
E_{N}(\mu_F^2,\mu_{0F}^2) &=& \left[{{\as(\mu_{0F}^2)}
\over{\as(\mu_F^2)}}\right]^{{{P_N^{(0)}}
\over{2\pi b_0}}} 
\exp\left\{ {{(\as(\mu_{0F}^2) - \as(\mu_F^2))}\over{4\pi^2 b_0}}
\left(P_N^{(1)} - {{2\pi b_1}\over{b_0} }  P_N^{(0)}\right) \right. \\
&+& \left. {\cal O}\left( \as^{n+2} \ln^n \frac{\mu_F^2}{\mu_{0F}^2} 
\right) \right\} \; , \nn
\eeeq
where the first factor on the right-hand side corresponds to the LO
approximation, the first term in the curly bracket gives the NLO correction,
and so forth.
The {\em leading} collinear logarithms $(\as \ln Q^2/m^2)^n$ of 
Eq.~(\ref{sighq})
are resummed by combining the LO expression of the evolution operator with
the LO evaluation of the partonic cross sections ${\hat \sigma}_a$ and of the
initial condition $D_{a/{\cal Q},\,N}^{{\rm ini}}$. The resummation
of the {\em next-to-leading} collinear terms $\as (\as \ln Q^2/m^2)^n$
requires the NLO evaluation of the evolution operator, of the partonic cross
sections and of the initial condition.

Although the perturbative fragmentation function is process-independent, it is
not unambiguously computable. More precisely, in Eq.~(\ref{sighq})
the separation between partonic cross sections and the perturbative 
fragmentation function depends on the factorization scheme. Since
massless partonic cross sections are usually evaluated in the \ms\ 
factorization scheme, in the following we always consider their definition
in this scheme. This (implicitly) fixes the factorization scheme for 
the perturbative fragmentation function.

Since the masses of the charm and bottom quarks are not very large,
these quarks typically undergo a non-perturbative fragmentation process rather
than perturbatively decay into lighter partons. In this non-perturbative
process,
hadronization takes place, eventually producing, for instance, an observable 
heavy meson $H$. Heavy-hadron fragmentation cross sections can still
be computed by using the factorization formula (\ref{sighq}), provided the
initial condition $D_{b/{\cal Q}}^{{\rm ini}}$ for the perturbative
fragmentation function is convoluted with a non-perturbative fragmentation 
distribution $D_{{\cal Q}/H}^{{\rm np}}$~\cite{Colangelo:1992kh,
Cacciari:1997du,Nason:2000zj}.
The most popular approach to describe $D_{{\cal Q}/H}^{{\rm np}}$ is the
phenomenological model by Peterson et al. \cite{Peterson:1983ak}.
Other approaches \cite{Jaffe:1994ie, Braaten:1995bz}, based on heavy-quark 
effective theory, are available.
The hadronization process can also be described by modelling non-perturbative
effects in the perturbative fragmentation function 
\cite{Dokshitzer:1996ev, Nason:1997pk}.
Recent studies \cite{shape} of non-perturbative effects may suggest other
descriptions of $D_{{\cal Q}/H}^{{\rm np}}$, based on a shape function
to be matched to the perturbative part.

At present the initial condition in Eq.~(\ref{dinfo}) is known up to NLO.
The calculation was first performed in Ref.~\cite{Mele:1991cw} by computing the
single inclusive cross section for heavy-quark production 
in \ee\ annihilation and subtracting the corresponding cross section in the
massless case. Using the \ms\ factorization scheme and limiting ourselves to 
the flavour non-singlet channel, the result is
\beq
\label{ininlo}
D^{{\rm ini}}(x,\as(\mu_0^2);\mu_0^2,\mu_{0F}^2,m^2)
\!= \delta(1-x) + \frac{\as(\mu_0^2)}{2 \pi} \;C_F  
\left[ \frac{1+x^2}{1-x} \left( \ln \frac{\mu_{0F}^2}{(1-x)^2 m^2} - 1 \right)
\right]_+ + {\cal O}(\as^2) ,
\eeq
where the plus-distribution is defined in Eq.~(\ref{+}). 

Considering the $N$ moments of Eq.~(\ref{ininlo}) in the large-$N$ limit,
we have
\beeq
\label{ininlon}
D_N^{{\rm ini}}(\as(\mu_0^2);\mu_0^2,\mu_{0F}^2,m^2) \!&=&\!\!
1 +  \frac{\as(\mu_0^2)}{\pi} \, C_F\; \left[ - \ln^2N + 
\left( \ln \frac{m^2}{\mu_{0F}^2} - 2\gamma_E + 1 \right) \ln N \right.  \\
\!&+&\!\!\left.  1 - \frac{\pi^2}{6} + \gamma_E - \gamma_E^2 + \left( \gamma_E 
- \frac{3}{4} \right) \ln \frac{m^2}{\mu_{0F}^2} +
{\cal O}\left(\frac{1}{N} \right) \right] + {\cal O}(\as^2) \;\;, \nn
\eeeq
This expression contains logarithmic contributions proportional to $\ln^2 N$
and $\ln N$. After having factorized and resummed large collinear terms,
$(\as \ln \mu_F^2/\mu_{0F}^2)^n$, by the evolution of the perturbative
fragmentation function, we still have to deal with the presence of $\ln N$
terms in the initial condition. These terms spoil the convergence of the
fixed-order perturbative expansion at large $N$ (or, equivalenty, at large
$x$), and have to be resummed to all perturbative orders. 
Note that these terms are present independently of the analogous terms
that have already been factorized at scale $\mu_F \sim Q$ and, possibly, 
resummed in the process-dependent partonic cross sections ${\hat \sigma}_a$ of 
Eq.~(\ref{sighq}). The $\ln N$ enhancement in Eq.~(\ref{ininlon}) is a 
process-independent Sudakov effect due to soft-gluon radiation from the heavy
quark at the scale $\mu_0 \sim m$. This effect can be quantitatively 
more important than that in the partonic cross sections, since it is controlled
by the coupling $\as(\mu_0^2)$ that is larger than  $\as(\mu_F^2)$.

In the following, we first discuss a general method to compute
the initial condition $D^{{\rm ini}}$ at any order in perturbation theory.
We explicitly apply the method at NLO, and we re-derive the result  in
Eq.~(\ref{ininlo});
we then perform all-order resummation of the large-$\ln N$ terms up
to NLL accuracy.

\subsection{Quasi-collinear factorization and the perturbative initial
condition}
\label{sec:qcolfac}

To understand the origin of the initial condition for the perturbative
fragmentation function, it is convenient to compare the heavy-quark cross
section $\sigma_{\cal Q}(x,Q;m)$ in Eq.~(\ref{sighq}) with the corresponding
perturbative cross section, $\sigma_q(x,Q;\ep)$, for the production of 
a massless quark $q$. The latter is collinearly divergent, and thus we
regularize the divergences by working in $d=4-2\ep$ space-time dimensions.
In these cross sections, both the massive and massless quarks are produced
at the large scale $Q$; they then perturbatively fragment by decreasing 
their momentum fraction $x$ and radiating some amount $\qt$ of transverse 
momentum. As long as $\qt$ is large, the massive quark behaves like a massless
one and, thus, $d\sigma_{\cal Q}/d\qt=d\sigma_q/d\qt$.
The difference between $\sigma_{\cal Q}$ and $\sigma_q$ is produced
by radiation at low transverse momentum, say $\q2t < \mu_{0F}^2$, where
$\mu_{0F}$ is an arbitrary scale such that $Q \gg \mu_{0F}$ (and 
$\mu_{0F} > m$ in the massive case). Since we are interested in the limit
$m/Q \ll 1$, we can choose a value of $\mu_{0F}$ that is parametrically very
small and exploit the universal factorization properties \cite{collpc}
of QCD radiation at low transverse momenta to write the $N$ moments
of the cross sections as
\beeq
\label{sigQ}
\sigma_{{\cal Q},N}(Q;m) &=& \sum_{a} {\widetilde \sigma}_{a,N}(Q;\mu_{0F})
\;{\widetilde D}_{a/{\cal Q},N}^{H}(\mu_{0F},m) \;+{\cal O}(m/Q) \;\;, \\
\label{sigq}
\sigma_{q,N}(Q;\ep) &=& \sum_{a} {\widetilde \sigma}_{a,N}(Q;\mu_{0F})
\;{\widetilde D}_{a/q,N}^{L}(\mu_{0F},\ep) \;+{\cal O}(\ep) \;\;. 
\eeeq
The factor ${\widetilde \sigma}$ on the right-hand side comes
from the high-$\qt$ region and therefore, in the limits $m \to 0$ and
$\ep \to 0$, it equally contributes to Eqs.~(\ref{sigQ}) and (\ref{sigq}).
The fragmentation contributions ${\widetilde D}_{a/{\cal Q},N}^{H}(\mu_{0F},m)$
and ${\widetilde D}_{a/q,N}^{L}(\mu_{0F},\ep)$, which come from the low-$\qt$ 
region ($\q2t < \mu_{0F}^2$), are different in the heavy-quark and 
massless-quark cases. These contributions, which are process-independent,
are divergent in the limits $m\to 0$ and $\ep \to 0$,
respectively. In the heavy-quark case, the divergences are avoided by keeping
$m$ finite, although neglecting corrections suppressed by powers of
$m/\mu_{0F}$ when $m/\mu_{0F} \to 0$. In the massless case, the divergences are
$\ep$-poles and have to be properly factorized to obtain the short-distance 
partonic cross section ${\hat \sigma}_{a,N}$
that appears in Eqs.~(\ref{fac}) and (\ref{sighq}). The latter is 
defined by\footnote{To simplify the notation, we set $\mu=\mu_F=\mu_{0F}$ in
Eq.~(\ref{sighq}).}
\beq
\label{sadef}
{\hat \sigma}_{a,N}(Q, \mu_{0F}) = \sum_b \sigma_{b,N}(Q;\ep) \;
\left[ {\bf \Gamma}^{(\msbar)}_N(\mu_{0F},\ep) \right]^{-1}_{ba} \;\;,
\eeq
where $\sigma_{b,N}(Q;\ep)$ is the dimensionally regularized
cross section in Eq.~(\ref{sigq}) and
${\bf \Gamma}$ is a matrix with respect to the flavour indices $a,b$
of the massless partons. Owing to the factorization theorem
of mass singularities \cite{book}, this matrix is process-independent (though
factorization-scheme-dependent) and contains all the $\ep$-poles that cancel 
the singularities of $\sigma_{b,N}(Q;\ep)$ in the limit $\ep \to 0$.
The explicit expression of ${\bf \Gamma}^{(\msbar)}$ as a function of the
anomalous dimensions $\gamma_{ab,N}(\as)$ in the \ms\ factorization scheme
is given in Ref.~\cite{cfp}.

Inserting Eq.~(\ref{sigq}) in Eq.~(\ref{sadef}),
we have ${\hat \sigma} = {\widetilde \sigma} {\widetilde D}^L \Gamma^{-1}$.
This expression for ${\hat \sigma}$ can be inserted in Eq.~(\ref{sighq}) 
to get $\sigma_{\cal Q} = {\widetilde \sigma} {\widetilde D}^L \Gamma^{-1}
D^{{\rm ini}}$. By comparison with Eq.~(\ref{sigQ}), we thus obtain the final 
result
\beq
\label{masterini}
\sum_{b,c} {\widetilde D}_{a/b,N}^{L}(\mu_{0F},\ep)
\left[ {\bf \Gamma}^{(\msbar)}_N(\mu_{0F},\ep) \right]^{-1}_{bc}
D_{c/{\cal Q},\,N}^{{\rm ini}}(\as(\mu_0^2);\mu_0^2,\mu_{0F}^2,m^2) 
= {\widetilde D}_{a/{\cal Q},N}^{H}(\mu_{0F},m) \;.
\eeq
Note that the high-$\qt$ and process-dependent contributions 
${\widetilde \sigma}_{a,N}$ of Eqs.~(\ref{sigQ}) and (\ref{sigq}) 
do not appear in Eq.~(\ref{masterini}). The master equation (\ref{masterini})
relates only process-independent contributions due to fragmentation at low
transverse momenta, and it can be used to explicitly compute the initial
condition $D_{a/{\cal Q},\,N}^{{\rm ini}}$. To this purpose we have to
evaluate ${\widetilde D}^H$ and ${\widetilde D}^L$, cancel the $\ep$-poles
in ${\widetilde D}^L$ by using the known expression of 
${\bf \Gamma}^{(\msbar)}$, and perform the limits $\ep \to 0$ and 
$m/\mu_{0F} \to 0$.

\begin{figure}[t]
\begin{center}
\begin{picture}(130,100)(0,0) 
\ArrowLine           ( -55,45)(18,80) 
\ArrowLine           ( -55,45)(20,65)
\GBox                ( -30,45)(20,47){0}
\ArrowLine           ( -55,45)( 15,25)
\Line                ( -55,45)( 12,15)
\ArrowLine           ( -55,45)( 8,5)
\BCirc               ( -55,45){25}
\ArrowLine           (  120,45)( 190,25)
\Line                (  120,45)( 187,15)
\ArrowLine           (  120,45)( 183,5)
\ArrowLine           ( 180,45)(235,80)
\ArrowLine           ( 180,45)(238,65)
\BCirc               ( 180,45){10}
\ArrowLine                ( 145,45)(170,45)
\GBox                ( 190,45)(240,47){0}
\BCirc               (  120,45){25}
\Text                ( -55,45)[1]{$M$}
\Text                ( 180,47)[1]{$P$}
\Text                ( 120,45)[1]{$M$}
\Text                ( 65,46)[1]{$\Rightarrow$}
\Text                ( 25,80)[l]{$X$}
\Text                ( 25,52)[l]{$p_1$}
\Text                ( 2,85)[l]{$q_i$}
\Text                ( 220,85)[l]{$q_i$}
\Text                ( 240,80)[l]{$X$}
\Text                ( 248,52)[l]{$p_1$}
\Text                ( 150,60)[l]{$p_1/z$}
\end{picture}
\caption{\label{factfig}\small Schematic picture of matrix element 
factorization in the collinear and quasi-collinear limits. The thicker
line denotes the particle (massless parton or heavy quark) that undergoes the
collinear or quasi-collinear splitting process.
}
\end{center}
\end{figure}
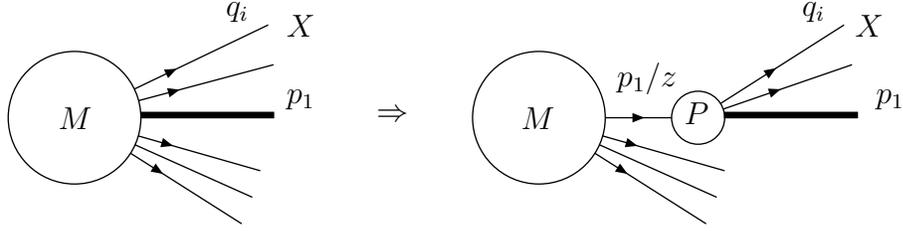

The factorization of the low-$\qt$ contributions ${\widetilde D}^H$ 
and ${\widetilde D}^L$ in Eqs.~(\ref{sigQ}) and (\ref{sigq}) follows
from the corresponding factorization formulae (Fig.~\ref{factfig}) of the 
QCD matrix elements (evaluated in physical gauges).
As is well known \cite{collpc, cfp, infrafac2, infrafac1}, in the case of
massless QCD the matrix element factorization for the splitting process
$a \to q(p_1) +X(\{q_i\})$ ($X$ denotes a set of partons with momenta
$q_i$ and total momentum $q = \sum_i q_i$) is controlled by the collinear 
limit.
As emphasized in Ref.~\cite{Catani:2001ef}, the dynamics of the splitting
processes of massive partons is described by analogous factorization 
formulae, which are obtained by generalizing the collinear limit to the {\em 
quasi-collinear} limit. Considering the splitting process
$a \to {\cal Q}(p_1) +X(\{q_i\})$, ${\cal Q}$ being a heavy quark of mass $m$
and momentum $p_1^\nu \;(p_1^2=m^2)$, we can describe its kinematics
in terms of the following Sudakov parametrization:
\beeq
p_1^\nu &=& z p^\nu - \qt^\nu + \frac{\q2t + (1-z^2) m^2}{z} \; 
\frac{n^\nu}{2p\cdot n} \;\; , \nn \\
\label{sudpar}
q^\nu &=& (1-z) p^\nu + \qt^\nu + \frac{q^2 + \q2t - (1-z)^2 m^2}{1-z} \;
\frac{n^\nu}{2p\cdot n} \;\; ,
\eeeq
where the momentum $p^\nu$ (with $p^2=m^2$) denotes the collinear (`forward')
direction,  and the light-like vector $n^\nu$ ($n^2=0$) denotes the
`backward' direction $(\qt \cdot p = \qt \cdot n = 0, \;
\q2t = - \qt^\nu q_{\perp \nu})$. The customary Sudakov parametrization
for the splitting process of a massless quark is recovered from 
Eq.~(\ref{sudpar}) by simply setting $m=0$. We can now consider the collinear
or quasi-collinear limits. In the massless case the collinear
region is defined by the limit $\qt \to 0$. In the heavy-quark case 
the quasi-collinear region is defined \cite{Catani:2001ef} by the limits
$\qt \to 0$ and $m \to 0$ at {\em fixed} ratio $m^2/\q2t$.
Note that the key difference between the collinear and quasi-collinear
limits is that the latter has to be performed by keeping the mass $m$ of the
same order as $\qt$ while $\qt \to 0$.

Performing the calculation of the collinear or quasi-collinear limits, 
a generic QCD matrix element 
$M(p_1,\{q_i\};\dots)$ factorizes in terms of a process-dependent
contribution (which is finite in the limiting region) and a universal AP
splitting function \cite{infrafac1, infrafac2, Catani:2001ef}. 
The fragmentation contributions 
${\widetilde D}^L$ and ${\widetilde D}^H$ in Eq.~(\ref{masterini}) are
obtained by integrating the AP splitting functions over the momenta $q_i$
of the partons involved in the corresponding splitting process. 

In this section we have so far outlined a general method to compute the
initial condition $D^{{\rm ini}}$ for heavy-quark fragmentation. As an
example, we now apply the method to the explicit calculation at NLO of the
flavour non-singlet component of $D^{{\rm ini}}$.

To this purpose, we have to consider the splitting process ${\cal Q} \to 
{\cal Q}(p_1) + g(q)$ at ${\cal O}(\as)$. Performing the collinear or
quasi-collinear limits, the matrix element $M(p_1,q;\dots)$ factorizes
as (see Fig.~\ref{factfig})
\beq
\label{melfac}
| M(p_1,q;\dots) |^2 \simeq | M(p_1/z;\dots) |^2 \;  
\frac{4 \pi \as}{p_1\cdot q} \; {\hat P} = | M(p_1/z;\dots) |^2 \;
8 \pi \as \;\frac{z(1-z)}{\q2t + (1-z)^2 m^2} \; {\hat P} \;, 
\eeq
where we have used the parametrization in Eq.~(\ref{sudpar}), the mass-shell
condition $q^2=0$ and the identity
\beq
2p_1\cdot q = 
\frac{\q2t + (1-z)^2 m^2}{z(1-z)} \;\;.
\eeq
In the massless case, ${\hat P}$ is the usual AP splitting function
in $d=4-2\ep$ dimensions,
\beq
\label{apcol}
{\hat P}_{qg}(z;\ep) = C_F \left[ \frac{1+z^2}{1-z} - \ep (1-z) \right]\, ,
\eeq
while its generalization to the heavy-quark case  
in $d=4$ dimensions is
\beq
\label{apqcol}
{\hat P}_{{\cal Q}g}(z;m^2/\q2t) = C_F \left[ \frac{1+z^2}{1-z} - 
\frac{m^2}{p_1\cdot q} \right]
= C_F \left[ \frac{1+z^2}{1-z} - \frac{2z(1-z)m^2}{\q2t + (1-z)^2 m^2} 
\right] \, .
\eeq
To extract the low-$\qt$ contributions
${\widetilde D}^L$ and ${\widetilde D}^H$ in Eq.~(\ref{masterini}), 
we consider the $N$ moments of the factorization formula 
(\ref{melfac}) and perform the integration of the AP splitting functions in
Eqs.~(\ref{apcol}) and (\ref{apqcol}) over the low transverse-momentum region,
$\q2t < \mu_{0F}^2$, of the radiated gluon.
The phase space for gluon radiation at fixed value of the longitudinal 
momentum $p_1\cdot n$ of the triggered quark is
\beq
d\Phi = (\mu_r)^{4-d} \;\frac{d^d q}{(2\pi)^{d-1}} \;\delta_+(q^2) = 
\frac{1}{16 \pi^2}
\;\frac{1}{\Gamma(1-\ep)} \;d\q2t \left( \frac{4 \pi \mu_r^2}{\q2t} 
\right)^\ep
\frac{d z}{z(1-z)} \;\Theta(z(1-z)) \; ,
\eeq
where $\mu_r$ is the dimensional-regularization scale. We thus obtain
\beeq
\!\!\!\!\!\!\!\!\! {\widetilde D}_{N}^{L}(\mu_{0F},\ep) &\!\!\!=&\!\!\!\! 1 +
\frac{\as}{2 \pi} \;\frac{1}{\Gamma(1-\ep)} \int_0^{\mu_{0F}^2} d\q2t
\;\left( \frac{4 \pi \mu_r^2}{\q2t} \right)^\ep \int_0^1 dz \;(z^{N-1} -1)
\;\frac{1}{\q2t} \; {\hat P}_{qg}(z;\ep) \nn \\
\label{m0case}
&\!\!\!=&\!\!\!\! 
1 - \frac{\as}{2 \pi} \;\frac{1}{\Gamma(1-\ep)}\;\frac{1}{\ep}
\;\left( \frac{4 \pi \mu_r^2}{\mu_{0F}^2} \right)^\ep
\int_0^1 dz \;(z^{N-1} -1) \;C_F \left[ \frac{1+z^2}{1-z} - \ep (1-z) 
\right] ,
\eeeq
and
\beeq
\label{masin}
&&\!\! \!\!\!\!\!\!\!\!{\widetilde D}_{N}^{H}(\mu_{0F},m) = 1 + 
\frac{\as}{2 \pi} \; \int_0^{\mu_{0F}^2} d\q2t
\; \int_0^1 dz \;(z^{N-1} -1)
\;\frac{1}{\q2t + (1-z)^2 m^2} \; {\hat P}_{{\cal Q}g}(z;m^2/\q2t) 
\\
&\!\!\!=\!\!\!&\! 1 + \frac{\as}{2 \pi} \;C_F \,\int_0^1 dz \;(z^{N-1} -1) 
\left[ \frac{1+z^2}{1-z} \ln \frac{\mu_{0F}^2 + (1-z)^2 m^2}{(1-z)^2 m^2}
- \frac{2z}{(1-z)} \frac{\mu_{0F}^2}{\mu_{0F}^2 + (1-z)^2 m^2} \right]
\nonumber \\
\label{masfin}
&\!\!\!=\!\!\!&\! 1 +  \frac{\as}{2 \pi} \;C_F \,\int_0^1 dz \;(z^{N-1} -1) 
\left[ \frac{1+z^2}{1-z} \ln \frac{\mu_{0F}^2}{(1-z)^2 m^2}
- \frac{2z}{(1-z)}  \right] + {\cal O}(m^2/\mu_{0F}^2) \, .
\eeeq
Note that the massless-quark calculation in Eq.~(\ref{m0case})
has consistently been performed by using dimensional regularization, while
in the heavy-quark calculation of Eq.~(\ref{masfin})
we have set $d=4$. The weight factor $(z^{N-1} -1)$ in Eqs.~(\ref{m0case})
and (\ref{masfin}) takes into account real (`$z^{N-1}$') and virtual
(`$-1$') gluon radiation. The real contribution derives directly from
Eq.~(\ref{melfac}). The virtual contribution has been included by simply
enforcing the constraint ${\widetilde D}_{N=1}^{L}={\widetilde D}_{N=1}^{H}
=1$, which follows from the conservation of the fermion number in the
non-singlet sector.

The initial condition $D_{N}^{{\rm ini}}$ is obtained from the master 
equation (\ref{masterini}), which in the non-singlet sector simplifies as
\beq
\label{masterns}
D_{N}^{{\rm ini}}(\as(\mu_0^2);\mu_0^2,\mu_{0F}^2,m^2) =
\frac{{\widetilde D}_{N}^{H}(\mu_{0F},m)}{{\widetilde D}_{N}^{L}(\mu_{0F},\ep)
\left[ {\Gamma}^{(\msbar)}_N(\mu_{0F},\ep) \right]^{-1}} \;\;,
\eeq
or, equivalently, at ${\cal O}(\as)$:
\beq
\label{dinipert}
D_{N}^{{\rm ini}}(\as(\mu_0^2);\mu_0^2,\mu_{0F}^2,m^2) = 1 +
\left\{ {\widetilde D}_{N}^{H}(\mu_{0F},m)
- {\widetilde D}_{N}^{L}(\mu_{0F},\ep)
\left[ \Gamma^{(\msbar)}_N(\mu_{0F},\ep) \right]^{-1} \right\}
+ {\cal O}(\as^2) \;\;.
\eeq
To evaluate the massless-quark contribution ${\widetilde D}^{L}
{\Gamma}^{-1}$, we use the 
${\cal O}(\as)$ expression of the collinear counterterm
$\Gamma^{(\msbar)}_N$ in the \ms\ factorization scheme \cite{cfp}:
\beq
\label{gmsbar}
\Gamma^{(\msbar)}_N(\mu_{0F},\ep) = 1 - \frac{\as}{2 \pi}
\;\frac{1}{\Gamma(1-\ep)}\;\frac{1}{\ep}
\;\left( \frac{4 \pi \mu_r^2}{\mu_{0F}^2} \right)^\ep
\int_0^1 dz \;(z^{N-1} -1) \;C_F \;\frac{1+z^2}{1-z}
+ {\cal O}(\as^2) \, .
\eeq
Using Eqs.~(\ref{m0case}) and (\ref{gmsbar}) we obtain
\beq
\label{m0fin}
{\widetilde D}_{N}^{L}(\mu_{0F},\ep)
\left[ \Gamma^{(\msbar)}_N(\mu_{0F},\ep) \right]^{-1} = 
1+ \frac{\as}{2 \pi} \;C_F \;\int_0^1 dz \;(z^{N-1} -1) \;\left[ 1-z \right]
+ {\cal O}(\ep) \, .
\eeq
Note that this expression contains no $\ep$ poles and the dependence
on the dimensional-regularization scale $\mu_r$ has consistently disappeared
in the limit $\ep \to 0$. Note also that the finite (when $\ep \to 0$)
term on the right-hand side {\em entirely} derives from the 
$\ep$-dependent part of the AP splitting function in Eq.~(\ref{apcol}).

Inserting Eq.~(\ref{m0fin}) in Eq.~(\ref{dinipert}) and using 
Eq.~(\ref{masfin}), we finally obtain the ${\cal O}(\as)$ contribution to 
the initial condition:
\beeq
D^{\rm ini}_{N}(\as(\mu_0^2);\mu_0^2,\mu_{0F}^2,m^2) &\!\!\!=&\!\!\! 1 + 
\frac{\as}{2 \pi} \;C_F \;\int_0^1 dz \;(z^{N-1} -1) 
\left[ \frac{1+z^2}{1-z} \left(\ln \frac{\mu_{0F}^2}{(1-z)^2 m^2} 
- 1\right)\right] \nn \\
&\!\!\!+&\!\!\! {\cal O}(\as^2) \;\;.
\eeeq
This result agrees with the known \cite{Mele:1991cw} expression in
Eq.~(\ref{ininlo}). Our calculation explicitly shows that the 
plus-distribution in Eq.~(\ref{ininlo}) receives two contributions: the term 
in the square bracket of Eq.~(\ref{masfin}) 
and that in the square bracket of Eq.~(\ref{m0fin}). The 
former is due to heavy-quark fragmentation at low $\qt$, while the latter
entirely depends on the prescription (dimensional regularization and
\ms\ factorization) used to handle the collinear divergences in the
fragmentation of the massless quark. A derivation of Eq.~(\ref{ininlo})
similar to ours is presented in Ref.~\cite{Keller:1999tf}.

\subsection{Soft-gluon resummation for the initial condition}
\label{sec:resHQ}

The ${\cal O}(\as)$ calculation presented in the second part of
Sect.~\ref{sec:qcolfac} can be extended to the flavour singlet components
of $D_{a/{\cal Q}}^{\rm ini}$. The method discussed in the first part
of Sect.~\ref{sec:qcolfac} can also be used to evaluate $D^{\rm ini}$
at ${\cal O}(\as^2)$ (and at higher orders), provided the calculation of the
collinear (massless) factorization formulae of 
Refs.~\cite{infrafac1, infrafac2} is generalized
to the quasi-collinear (heavy-quark) case. In this section we use the master
equation (\ref{masterini}) (or Eq.~(\ref{masterns})) to perform all-order 
resummation of the soft-gluon contributions to $D_{N}^{\rm ini}$ at large $N$
(or, equivalently, at large $x$). We limit ourselves to considering the 
non-singlet component, since the flavour-singlet contributions are suppressed
by a relative factor of ${\cal O}(1/N)$ when $N \to \infty$.

At high perturbative orders the fragmentation factors
${\widetilde D}_{N}^{H}$ and ${\widetilde D}_{N}^{L}$
receive contributions from multiple radiation of partons with momenta $q_i$.
All-order resummation of these contributions is achieved  as in the case of
the customary AP evolution in transverse momentum. The region of ordered
transverse momenta, $\mu_{0F}^2 > \dots > {\bf q}^2_{\perp \,i} 
> {\bf q}^2_{\perp \,j} > \dots > 0$, leads to the exponentiation of the
lowest-order AP splitting kernel. Then, the subregions where two or more
transverse momenta are of the same order (${\bf q}^2_{\perp \,i} \sim \dots
\sim {\bf q}^2_{\perp \,j}$) lead to perturbative corrections to the
exponentiated kernel. In particular, in the case of soft-gluon radiation
the inclusive correction to the lowest-order AP kernel can be taken into
account by the simple replacement \cite{ct2, cmw}
\beq
\label{asrep}
\as \to \as(\q2t) \left[ 1+ \frac{\as(\q2t)}{2\pi} \left( K + {\cal O}(\ep)
\right) + {\cal O}(\as^2) \right] \;\;,
\eeq
where the coefficient $K$ is given in Eq.~(\ref{kcoef}), and the term
${\cal O}(\as^2)$ only contributes beyond NLL accuracy at large $N$.
This resummation procedure applies to all the factors, 
${\widetilde D}_{N}^{L}, \Gamma^{(\msbar)}_N$ and ${\widetilde D}_{N}^{H}$,
in Eq.~(\ref{masterns}).

We first consider the case of massless-quark fragmentation. We note from 
Eq.~(\ref{m0fin}) that the factor ${\widetilde D}_{N}^{L}
[\Gamma^{(\msbar)}_N]^{-1}$ does not show any $\ln N$ enhancement at 
${\cal O}(\as)$. Soft-gluon radiation produces $\ln N$ corrections in both  
Eqs.~(\ref{m0case}) and (\ref{gmsbar}), but they cancel each other.
Owing to Eq.~(\ref{asrep}), this cancellation mechanism is valid up to
NLL order in the \ms\ factorization scheme, since both factors
${\widetilde D}_{N}^{L}(\mu_{0F},\ep)$ and 
${\bf \Gamma}^{(\msbar)}_N(\mu_{0F},\ep)$ include soft-gluon emission
up to the transverse-momentum scale $\q2t = \mu_{0F}^2$. We thus have
\beq
\label{m0all}
{\widetilde D}_{N}^{L}(\mu_{0F},\ep)
\left[ \Gamma^{(\msbar)}_N(\mu_{0F},\ep) \right]^{-1} = 
1+ {\cal O}\left(\as (\as \ln N)^k\right) \, .
\eeq

In the heavy-quark case, we first exponentiate Eq.~(\ref{masin}) and then 
perform the replacement of Eq.~(\ref{asrep}). We obtain
\beeq
\ln {\widetilde D}_{N}^{H}(\mu_{0F},m) &=& 
\int_0^1 dz \;(z^{N-1} -1) \;\int_0^{\mu_{0F}^2} 
\frac{d\q2t}{\q2t + (1-z)^2 m^2} \nn \\
&\cdot& 
\frac{\as(\q2t)}{2\pi} \left[ 1+ \frac{\as(\q2t)}{2\pi} K \right]
\; {\hat P}_{{\cal Q}g}(z;m^2/\q2t) 
+ {\cal O}\left(\as (\as \ln N)^k\right) \;\;.
\eeeq
The right-hand side can be further
simplified by neglecting contributions of ${\cal O}(m^2/\mu_{0F}^2)$ and
terms beyond NLL order at large $N$. To this purpose, we note that 
${\hat P}_{{\cal Q}g}$ contains two different contributions. In the limit
$m \to 0$, the $\q2t$-integrals of the first and second contributions in the 
square bracket of Eq.~(\ref{apqcol}) are respectively logarithmically and 
linearly divergent at small $\q2t$. To the required accuracy, the 
logarithmically-divergent and linearly-divergent integrals are dominated by 
the transverse-momentum regions 
$\mu_{0F}^2 > \q2t > (1-z)^2 m^2$ and $\q2t \simeq (1-z)^2 m^2$, respectively.
We thus obtain
\beeq
\ln {\widetilde D}_{N}^{H}(\mu_{0F},m) &=&
\int_0^1 dz \;\frac{z^{N-1} -1}{1-z} \left\{ \int_{(1-z)^2 m^2}^{\mu_{0F}^2} 
\frac{d\q2t}{\q2t} 
\; \frac{\as(\q2t)}{\pi} \left[ 1+ \frac{\as(\q2t)}{2\pi} K \right] \right. 
\nn \\
\label{dHres}
&-& \left. \frac{C_F}{\pi} \, \as((1-z)^2 m^2)
\right\} + {\cal O}\left(\as (\as \ln N)^k\right) \;.
\eeeq
The expression in the curly bracket clearly shows the presence of two
characteristic transverse-momentum scales: the factorization scale
$\mu_{0F}^2$ and the heavy-quark scale $(1-z)^2 m^2$. The latter is related to
the angular size $\theta_0=m/E_{\cal Q}$ ($E_{\cal Q}$ is the energy of the
heavy quark) of the `dead cone' \cite{Dokshitzer:1996ev} for bremsstrahlung 
off a massive particle. Writing the transverse momentum as 
$\q2t \simeq (1-z)^2 E_{\cal Q}^2 \theta^2$, where $(1-z) E_{\cal Q}$ is the
energy of the radiated gluon and $\theta$ is its emission angle, the first 
term in the curly bracket of Eq.~(\ref{dHres}) describes soft and collinear
radiation outside the dead cone ($\theta^2 > \theta_0^2$), while the second
term is related to soft radiation near the dead-cone boundary 
($\theta^2 \simeq \theta_0^2$). 

Inserting Eq.~(\ref{m0all}) in  Eq.~(\ref{masterns}), we obtain the following 
resummed expression for the initial condition of the heavy-quark fragmentation
function:
\beq
\label{nllhres}
\ln D^{\rm ini}_{N}(\as(\mu_0^2);\mu_0^2,\mu_{0F}^2,m^2) =
\ln {\widetilde D}_{N}^{H}(\mu_{0F},m) + 
{\cal O}\left(\as (\as \ln N)^k\right) \;\;,
\eeq
where $\ln {\widetilde D}_{N}^{H}$ is given in Eq.~(\ref{dHres}) up
to NLL accuracy. Moreover, the derivation of Eqs.~(\ref{m0all}) and
(\ref{dHres}) and the dependence on the factorization scale $\mu_{0F}$
of the perturbative fragmentation function in Eq.~(\ref{pertev})
suggest a generalization of the NLL result in Eq.~(\ref{nllhres})
to any logarithmic order as follows:
\beeq
\ln D_N^{{\rm ini}}(\as(\mu_0^2);\mu_0^2,\mu_{0F}^2,m^2) &=&
\int_0^1 dz \;\frac{z^{N-1} -1}{1-z} \; \Bigl\{ \int_{(1-z)^2 m^2}^{\mu_{0F}^2} 
\frac{d\q2t}{\q2t} \;A[\as(\q2t)] \Bigr. \nn\\
\label{alllogHQ}
&+& \Bigl. H[\as((1-z)^2 m^2)]
\Bigr\} +{\cal O}(1/N) \;\;.
\eeeq
The function $A(\as)$, whose perturbative expansion is given in 
Eq.~(\ref{afun}), is the process-independent function that controls the 
large-$N$ behaviour of the flavour non-singlet anomalous dimensions 
(see Eq.~(\ref{andim})). The function $H(\as)$,
\beq
H(\as)=\sum_{n=1}^{\infty} \left( \asp \right)^n \;H^{(n)} \;,
\eeq
is strictly related to soft radiation off a heavy quark, and its
first-order coefficient is
\beq
\label{h1}
H^{(1)} = - \,C_F \, .
\eeq
The coefficient $A^{(1)}$ controls the resummation of the LL terms
$\as^n \ln^{n+1}N$ in Eq.~(\ref{alllogHQ}).
The coefficients $A^{(2)}$ and $H^{(1)}$ give the NLL terms $\as^n \ln^{n}N$,
and so forth. At LL accuracy, Eq.~(\ref{alllogHQ}) agrees with the resummed
calculation of Ref.~\cite{Mele:1991cw}. Knowing the coefficients $A^{(2)}$ 
and $H^{(1)}$ in Eqs.~(\ref{a1a2}) and (\ref{h1}), we have explicitly 
extended the result of Ref.~\cite{Mele:1991cw} up to NLL accuracy.

To quantitatively study the effect of soft-gluon resummation, we proceed as in
Sect.~\ref{sec:numL}. The Sudakov-resummed part $D_N^{{\rm ini}, \,S}$ of the 
initial condition is written as
\beeq
D_N^{{\rm ini}, \,S}(\as(\mu_0^2);\mu_0^2,\mu_{0F}^2,m^2) &\!\!\!=& \!\!\! 
\left\{ 1 + {{\as(\mu_0^2) C_F}\over{\pi}}\left[
-\frac{\pi^2}{6} + 1 - \gamma_E^2 + \gamma_E + \left(\frac{3}{4} -
\gamma_E\right) \ln\frac{\mu_{0F}^2}{m^2}\right]\right\}\nonumber\\
\label{sudHQini}
&\!\!\!\times&\!\!\! \exp \Bigl[ \ln N \;g^{(1)}_{\rm ini}(\lambda_0) + 
g^{(2)}_{\rm ini}(\lambda_0,m^2/\mu_0^2;m^2/\mu_{0F}^2) 
\Bigr]\, ,
\eeeq 
where
\beq
\lambda_0 = b_0 \as(\mu_0^2) \ln N\, ,
\eeq
and we have explicitly introduced the renormalization scale $\mu_0$, which,
in general, is different from the factorization scale $\mu_{0F}$.
The LL and NLL functions $g^{(1)}_{\rm ini}$ and $g^{(2)}_{\rm ini}$ are
obtained by performing the $z$ and $\q2t$ integrations of Eqs.~(\ref{alllogHQ})
up to NLL accuracy:
\beeq
\label{g1funin} 
g^{(1)}_{\rm ini}(\lambda_0) &=& 
-\frac{A^{(1)}}{2\pi b_0 \lambda_0} \;
\bigl[ 2\lambda_0 + (1-2\lambda_0) \ln (1-2\lambda_0) \bigr] \;,\\
\label{g2funin}
g^{(2)}_{\rm ini}(\lambda_0,m^2/\mu_0^2;m^2/\mu_{0F}^2) &=&
\frac{A^{(1)}}{2 \pi b_0}\left(\ln \frac{\mu_{0F}^2}{m^2} + 2\gamma_E\right) 
\ln(1-2\lambda_0)\nn\\
&&-\frac{A^{(1)}  b_1}{4 \pi b_0^3}
\left[ 4\lambda_0 + 2 \ln (1-2\lambda_0) +  \ln^2 (1-2\lambda_0) \right]
\nn \\
&&+ \frac{1}{2\pi b_0} \left[2\lambda_0 + \ln (1-2\lambda_0) \right]
\left(\frac{A^{(2)}}{\pi b_0} + A^{(1)}\ln\frac{\mu_{0}^2}{\mu_{0F}^2}\right)
\nn\\
&&+ \frac{H^{(1)}}{2 \pi b_0} \ln (1-2\lambda_0) \; .
\eeeq
Analogously to Eq.~(\ref{cneeres}), the terms in the curly bracket of 
Eq.~(\ref{sudHQini})
are the constant (when $N \to \infty$) contributions to the NLO initial
condition in Eq.~(\ref{ininlon}).

The final NLO+NLL resummed expression for the $N$ moments of the (non-singlet)
initial condition is then given by
\beeq
\label{resfin-dini}
D_N^{\rm ini (res)}(\as(\mu_0^2);\mu_0^2,\mu_{0F}^2,m^2)
&=& D_N^{{\rm ini}, S}(\as(\mu_0^2);\mu_0^2,\mu_{0F}^2,m^2)
- \left[ D_N^{{\rm ini}, S}(\as(\mu_0^2);\mu_0^2,\mu_{0F}^2,m^2) \right]_{\as} 
\nn \\
&+& \left[ D_N^{{\rm ini}}(\as(\mu_0^2);\mu_0^2,\mu_{0F}^2,m^2)
\right]_{\as}\;\;,
\eeeq
where $[ D_N^{{\rm ini}} ]_{\as}$ are the $N$ moments of the ${\cal O}(\as)$
result in Eq.~(\ref{ininlo}),
$D_N^{{\rm ini}, S}$ is given in Eq.~(\ref{sudHQini}) and 
$[ D_N^{{\rm ini}, S} ]_{\as}$ is its perturbative truncation
at ${\cal O}(\as)$.

Note that, analogously to Eq.~(\ref{cneeres}), the Sudakov-resummed part 
$D_N^{{\rm ini}, \,S}$ of the heavy-quark initial condition 
has cut singularities in the complex variable $N$. In the heavy-quark case
the singularities start at the branch-point $N=\exp\{1/2b_0 \as(\mu_0^2)\}$
(i.e. at $\lambda_0=1/2$ in Eqs.~(\ref{g1funin}) and (\ref{g2funin})). They
signal the dominance of non-perturbative effects at very large values of $N$
(or very large $x$). As in the massless case, we do not explicitly include any 
non-perturbative contributions in our calculations for heavy-quark
fragmentation, and we simply apply the Minimal Prescription of 
Ref.~\cite{Catani:1996yz} when performing numerical inversions of 
Mellin moments to $x$ space.

\begin{figure}[t]
\begin{center}
\epsfig{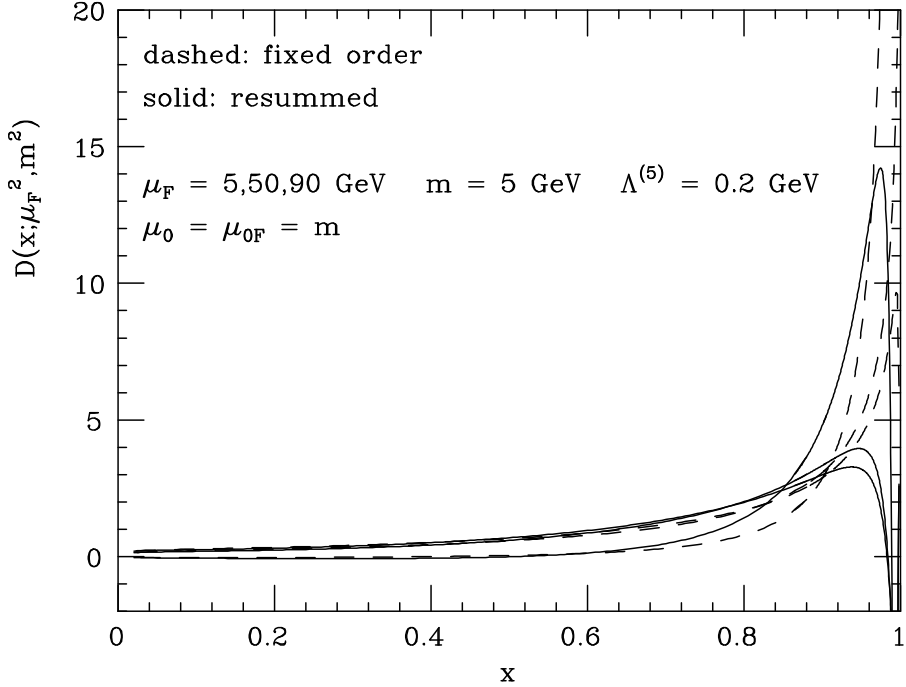}\hfill
\caption{\label{light}\small Effect of Sudakov resummation for the initial 
condition of the heavy-quark fragmentation function, 
evolved up to three different scales $\mu_F = 5, 50, 90$~GeV.}
\end{center}
\end{figure}

The NLO and NLO+NLL calculations are compared in Fig.~\ref{light}. In both
calculations we consider the perturbative fragmentation function 
$D(x;\mu_F^2,m^2)$, which is obtained from the corresponding initial condition 
$D^{\rm ini}$ at the scale $\mu_{0F}=\mu_0=m$ 
(the dependence on the scales $\mu_{0F}$ and $\mu_0$ is studied in 
Sect.~\ref{sec:eeHQ}) by AP evolution (see Eqs.~(\ref{pertev}) and
(\ref{nloeo})) up to the scale $\mu_F$. We can see that the NLO+NLL resummed
fragmentation function is much softer than the NLO fragmentation function.
Moreover, in both cases, the AP evolution reduces the heavy-quark momentum,
and the fragmentation function is softened.

A typical feature of all-order soft-gluon resummation is the presence of
a characteristic maximum (Sudakov peak) \cite{Dokshitzer:1996ev} at $x=x_{\rm
peak}$ in the NLO+NLL fragmentation function. The position of the peak is at
$1-x_{\rm peak} \sim (\Lambda/m)^c$, where the power $c=1-\exp\{-\pi
b_0/A^{(1)}\}$ is slightly smaller than unity.

Figure \ref{light} also shows that the fragmentation function becomes
negative when $x$ is very close to $x=1$. This occurs both in the NLO+NLL
calculation and in the NLO one (see also Fig.~\ref{scales-x}), although
in the NLO case the effect is not very evident from Fig.~\ref{light} because
it is due to a negative $\delta$-function contribution concentrated near $x=1$
(see Eq.~(\ref{ininlo})). In both cases the negative behaviour is a consequence
of the approximate character of the calculations for the initial condition
and, therefore, it is unphysical. Note, however, that the origin of this 
behaviour is different in 
the two cases. In the NLO case the negative behaviour is due to the presence 
of large, unresummed double-logarithmic terms $\as\ln^2(1-x)$. 
In the NLO+NLL case this pathological behaviour is cured by soft-gluon 
resummation, but the sensitivity to non-perturbative QCD phenomena still drives
the fragmentation function negative. As a matter of fact, if we had 
considered soft-photon resummation for the fragmentation of a massive lepton in
QED, the ensuing NLO+NLL fragmentation function would have been positive.
In QCD, instead, the Landau singularity of the perturbative coupling at small
transverse momenta leads to a branch-point in the resummed expression 
(\ref{sudHQini}). The branch-point occurs when $2b_0 \as(m^2) \ln N = 1$ 
in the $N$-plane and produces the negative behaviour of $D(x;\mu_F^2,m^2)$
at $1-x \sim \Lambda/m$. At such large values of $x$, non-perturbative
contributions have to properly be included in the evaluation of the heavy-quark
fragmentation function \cite{Dokshitzer:1996ev, Nason:1997pk}.

Soft-gluon resummation therefore suggests that non-perturbative
fragmentation phenomena become dominant when $1-x \sim \Lambda/m$. This
suggestion is consistent with the findings of the approach
\cite{Jaffe:1994ie, Braaten:1995bz} based on heavy-quark effective
theory. However, the expectation of the authors of
Ref.~\cite{Jaffe:1994ie} that Sudakov effects are large only when  $1-x
\sim (\Lambda/m)^2$, and likely to be less  important than
non-perturbative effects, is not correct. In fact, the resummed
calculation shows that the Sudakov effects are large at smaller values
of $x$. In particular, the Sudakov peak is placed at   $1-x_{\rm peak}
\sim (\Lambda/m)^c > \Lambda/m$. Around the region of the  Sudakov
peak, perturbative soft-gluon resummation is (at least) as important 
as non-perturbative effects.

The results presented in this section regard the process-independent 
fragmentation function
$D(x;\mu_F^2,m^2)$. This universal (though factorization-scheme-dependent) 
heavy-quark distribution enters the perturbative QCD predictions
of any process-dependent cross sections for heavy-quark fragmentation 
(see Eq.~(\ref{sighq})). In the next section, we explicitly consider  
heavy-quark fragmentation in \ee\ annihilation.

\subsection{Single-inclusive heavy-quark distribution in \ee\ annihilation}
\label{sec:eeHQ}

The most accurate data on the fragmentation of bottom and charm quarks
come from \ee\ annihilation experiments. These data can be used to test
perturbative QCD predictions and to extract information on the non-perturbative
contribution to heavy-quark fragmentation.

We consider the inclusive production of a single heavy meson $H$ in 
$e^+e^-$ collisions,
\beq
e^+ + e^- \to V(Q) \to H(p) + X\, ,
\eeq
in the kinematical region where the centre-of-mass
energy $Q$ of the collision is much larger than the heavy-quark (heavy-meson)
mass $m$. This process is completely analogous to the process in 
Eq.~(\ref{gammapx}), apart from the replacement of the observed light-flavoured
hadron $h(p)$ by the heavy-flavoured hadron $H(p)$. 

We study the inclusive cross
section $d\sigma/dx$ with respect to the energy fraction $x=2p\cdot Q/Q^2$
of the heavy-flavoured hadron, and we define the inclusive distribution
\beq
\label{indistdef}
{\cal D}(x;Q^2,m^2) \equiv \frac{1}{\sigma_{\rm tot}}\frac{d\sigma}{dx}\, ,
\eeq
where $\sigma_{\rm tot}$ is the total hadronic cross section
in \ee\ annihilation.
Since we are mainly interested in the large-$x$
behaviour, we consider only the flavour non-singlet contribution to the cross
section. Using the perturbative fragmentation function formalism of
Eqs.~(\ref{sighq}) and (\ref{pertev}), we write the $N$ moments of the 
inclusive distribution as
\beq
\label{dnHQ}
{\cal D}_N(Q^2,m^2) = \frac{\sigma^{(LO)}}{\sigma_{\rm tot}} \;
C_N^{(e^+e^-)}(\as(\mu^2);Q^2,\mu^2,\mu_{F}^2) \; E_{N}(\mu_F^2,\mu_{0F}^2)
\;D_N^{\rm ini}(\as(\mu_0^2);\mu_0^2,\mu_{0F}^2,m^2) \;\;,
\eeq
where the ratio $\sigma^{(LO)}/\sigma_{\rm tot}$ is given in Eq.~(\ref{stot}),
$C_N^{(e^+e^-)}$ is the \ms\ coefficient function in Eq.~(\ref{dshad}),
$E_{N}$ is the AP evolution operator in Eq.~(\ref{nloeo}),
and $D_N^{\rm ini}$ is the perturbative initial condition for the heavy-quark
fragmentation function. To be precise, the right-hand side of Eq.~(\ref{dnHQ})
should contain an additional factor $D_{{\cal Q}/H, N}^{\rm np}$, describing 
the non-perturbative fragmentation of the heavy quark ${\cal Q}$ into the 
observed heavy meson $H$. Since we are mainly interested in the perturbative 
contributions to the inclusive distribution, we do not include this factor.

In the following we compare fixed-order and resummed calculations for the
inclusive distribution in Eq.~(\ref{indistdef}). Analogously to the comparisons
in Sects.~\ref{sec:numL} and \ref{sec:resHQ}, the fixed-order calculation
uses the NLO expressions for $C_N^{(e^+e^-)}$ \cite{aemp, Nason:1994xx} and 
$D_N^{\rm ini}$ \cite{Mele:1991cw}, while the resummed calculation is obtained 
by using the expressions in Eqs.~(\ref{resfin}) and (\ref{resfin-dini}), 
which include the full NLO result and resum soft-gluon effects beyond 
${\cal O}(\as)$ to NLL accuracy.

We recall that soft-gluon resummation for the \ee\ inclusive distribution
${\cal D}$ was first considered in Ref.~\cite{Dokshitzer:1996ev}.
Comparing our resummed expression with the NLL results of 
Ref.~\cite{Dokshitzer:1996ev} in the ultrarelativistic limit $Q \gg m$,
we find complete agreement. This should be regarded as a consistency check
of our results. 
However, the authors of Ref.~\cite{Dokshitzer:1996ev} 
did not use the fragmentation function formalism.
Therefore, resummed formulae for the (process-dependent) massless coefficient 
function  $C_N^{(e^+e^-)}$ and the (process-independent) heavy-quark initial 
condition $D_N^{\rm ini}$ cannot be extracted separately from 
Ref.~\cite{Dokshitzer:1996ev}.

Since the most recent and accurate data on heavy-quark production in \ee\
annihilation come from $b$-quark fragmentation at LEP and SLC, in our 
numerical study 
we choose the centre-of-mass energy $Q = 90$~GeV, and the heavy-quark mass 
$m = 5$~GeV. 

\begin{figure}[t]
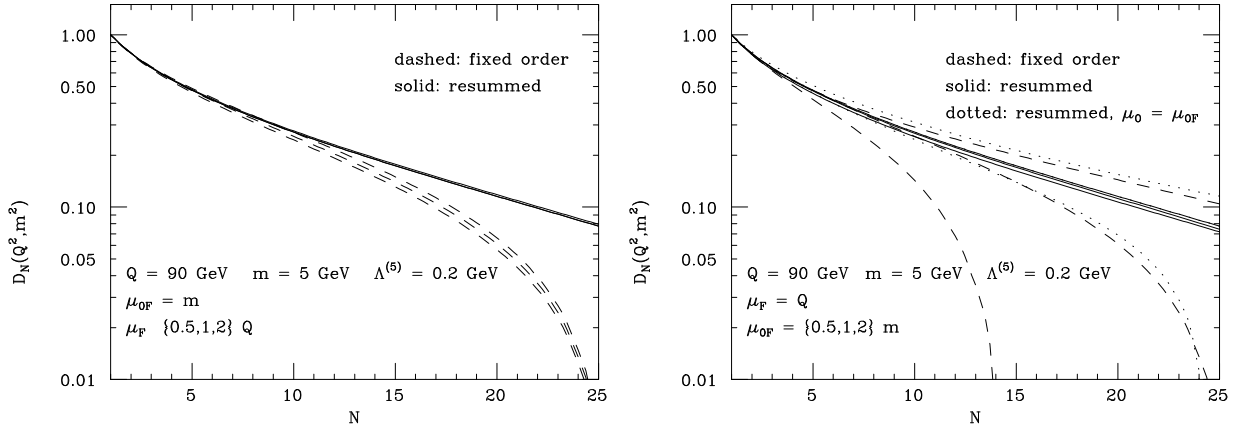

\begin{center}
\epsfig{file=mu-n.ps,width=8cm}\hfill
\epsfig{file=mu0-n.ps,width=8cm}
\caption{\label{scales-n}\small 
Dependence of the moments 
${\cal D}_N$ on the variations of the factorization scales $\mu_{F}$ and 
$\mu_{0F}$. The renormalization scales $\mu$ and $\mu_0$ are kept fixed at $Q$ 
and $m$ respectively, except for the dotted lines in the right-hand plot, 
where $\mu_0 =\mu_{0F}$.}
\end{center}
\end{figure}

\begin{figure}[t]
\begin{center}
\epsfig{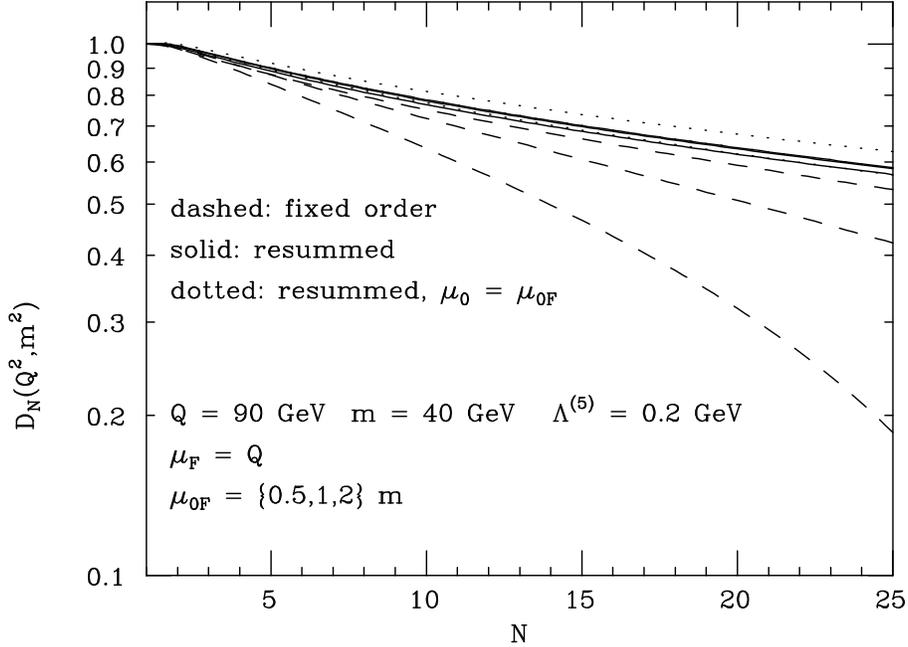}\hfill
\caption{\label{scales-n-m40}\small 
Dependence of the moments 
${\cal D}_N$ on the variation of the factorization and renormalization scales 
$\mu_{0F}$ and $\mu_0$ for a fictitious heavy quark of mass $m=40$ GeV.}
\end{center}
\end{figure}

We first consider the fixed-order and resummed calculations for
the $N$ moments of the heavy-quark distribution ${\cal D}(x;Q^2,m^2)$.
Figure~\ref{scales-n} shows the effect of varying the factorization scales 
$\mu_F$ and $\mu_{0F}$ in the ranges $\{Q/2,2Q\}$ and $\{m/2,2m\}$,
respectively.
The left panel shows the
effect of varying $\mu_F$, the right one that of varying $\mu_{0F}$. 
We use the customary practice of performing scale variations to estimate
(a lower bound on) the theoretical uncertainty due to uncalculated 
perturbative terms of higher order.
From the plot on the left-hand side, we observe that the 
`theoretical uncertainty' on the large-$N$ moments due to $\mu_F$ variations 
is smaller when resummation is performed. This effect derives from the 
resummation in the \ee\ coefficient function $C_N^{(e^+e^-)}$ of 
Eq.~(\ref{dnHQ}), and thus it is very similar to the effect already noticed
in Sect.~\ref{sec:numL} for the fragmentation spectrum of light hadrons 
(see, e.g., Fig.~\ref{fig:light}).  However, unlike those in Fig.~\ref{fig:light}, 
the NLO and NLO+NLL results in Fig.~\ref{scales-n} are very different,
well beyond the band due to $\mu_F$ variations. The difference is due to the
large effect produced by the resummation of the initial condition
$D_N^{\rm ini}$. Variations of the factorization scale $\mu_{0F}$ 
for the initial condition are considered in the plot on the right-hand side,
where we can see a very remarkable reduction in the scale dependence of the
resummed calculation. Since $\mu_{0F}$ is an auxiliary scale, introduced to 
perform the resummation of the large collinear terms $\as \ln Q^2/m^2$, the 
reduced scale dependence of the NLO+NLL calculation implies that Sudakov
resummation has highly improved collinear resummation in the large-$N$ region.
If we also vary the renormalization scale $\mu_0$ (dotted lines in the
right-hand 
plot), we observe an increased 'theoretical uncertainty' of the resummed
calculation. 
This is mainly because, in the calculation of the initial 
condition, we are using perturbation theory down to a fairly small scale
$\mu_0$, namely $\mu_0 \sim m \sim 5$~GeV.
To show how a limiting factor to the theoretical accuracy of the calculation
is the low scale set by the bottom mass, we consider a plot analogous to the 
right one in Fig.~\ref{scales-n}, but for a fictitious heavy quark of mass 
$m=40$~GeV. All other parameters remain identical. In Fig.~\ref{scales-n-m40}, 
we can clearly
see how the `theoretical uncertainty' is now greatly reduced by resummation, 
even when varying also the renormalization scale $\mu_0$.

\begin{figure}[t]
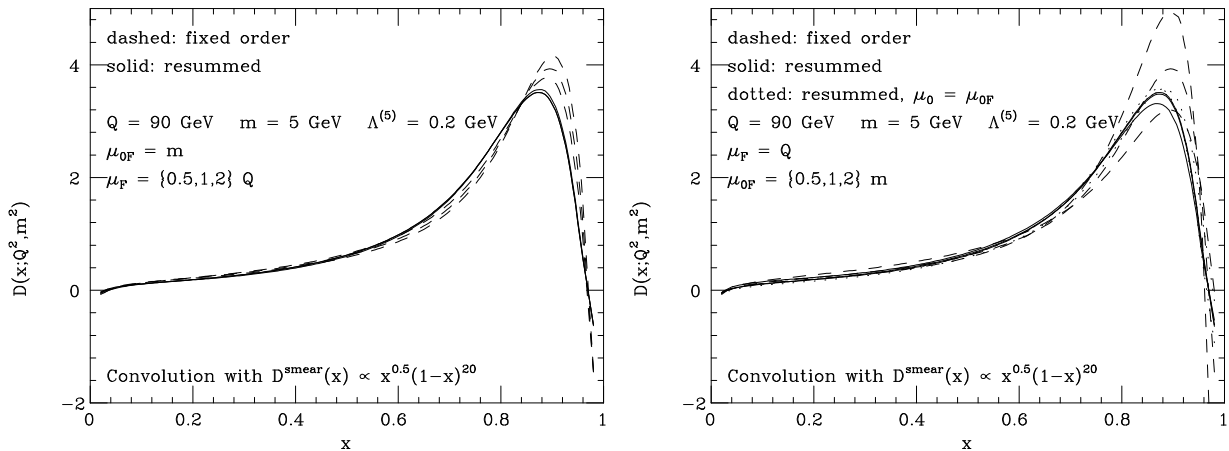

\begin{center}
\epsfig{file=mu-x.ps,width=8cm}\hfill
\epsfig{file=mu0-x.ps,width=8cm}
\caption{\label{scales-x}\small 
Dependence of the heavy-quark distribution
${\cal D}(x)$ on the variations of the factorization scales $\mu_F$ (left) and
$\mu_{0F}$ (right). Convolution with a smearing
function $D^{\rm smear}(x) \propto x^{0.5} (1-x)^{20}$, normalized to
have $D^{\rm smear}_{N=1}=1$, has been included for 
better visibility. The dotted lines in the right-hand plot are obtained by varying
also the renormalization scale $\mu_0$ ($\mu_0 =\mu_{0F}= \{0.5, 2\}$).}
\end{center}
\end{figure}

To consider also the $x$-space distribution, we perform a 
numerical inverse Mellin transformation of Eq.~(\ref{dnHQ}) by 
using the Minimal Prescription of Ref.~\cite{Catani:1996yz}, as already 
described in Sects.~\ref{sec:numL} and \ref{sec:resHQ}.
Figure~\ref{scales-x} shows again the effect of varying the factorization 
scales, this time on the $x$-space inclusive distribution 
${\cal D}(x;Q^2,m^2)$. A smearing
function $D^{\rm smear}(x) \propto x^{0.5} (1-x)^{20}$, properly 
normalized so that $\int^1_0 dx\, D^{\rm smear}(x) = 1$, has been 
convoluted with the purely perturbative results for better visibility. 
As in the case of the $N$ moments (Fig.~\ref{scales-n}), inclusion of Sudakov
resummation greatly reduces the dependence on  
factorization-scale variations: the shape of the perturbative contribution
to the inclusive distribution can now
be more reliably predicted. The improvement also looks robust with respect to
variations of the renormalization scale $\mu_0$ (dotted lines in the
right-hand plot). We can also observe that the unphysical behaviour (its origin
has been discussed at the end of Sect.~\ref{sec:resHQ})
very close to $x=1$, where the heavy-quark distribution turns negative,
is mitigated by the inclusion of the resummed coefficient function and 
initial condition.

As shown by the numerical results presented in this section (and in 
Sect.~\ref{sec:resHQ}), Sudakov resummation has important effects
on heavy-quark fragmentation. In particular, the reduced scale-dependence of 
the resummed calculation permits a better control of the non-perturbative 
contributions to the heavy-quark fragmentation function.
Most of these effects follow from the resummation
in the perturbative initial condition and are, therefore, process-independent.
For practical phenomenological purposes, one may thus think of evaluating
the initial condition by considering only fixed-order perturbative 
contributions, without explicitly resumming Sudakov-enhanced contributions.
The latter would then `effectively' be reabsorbed in the definition of the
non-perturbative component of the heavy-quark fragmentation function.
However, besides it being poorly justified on theoretical grounds, such 
an approach
would also lead to much larger uncertainties due to the much stronger
scale dependence of the fixed-order calculation.

\section{Summary}
\label{sec:sum}

In this paper we have performed soft-gluon resummation with 
NLL accuracy for fragmentation processes of light and heavy
quarks (hadrons) at high momentum fraction.

In the light-quark case, the fragmentation cross sections (see Eq.~(\ref{fac}))
are obtained
by convoluting process-dependent coefficient functions with the 
non-perturbative and process-independent fragmentation functions.
We have provided the explicit resummed expressions 
for the non-singlet \ms\ coefficient functions of the one-particle and
two-particle inclusive distributions in \ee\ collisions and the single-particle
inclusive cross section in DIS. We have studied in detail the \ee\
one-particle distribution by matching (see Eq.~(\ref{resfin}))
the NLL resummed part with the complete calculation at NLO.
From the mumerical comparison of our result with the NLO approximation, 
we observe that resummation stabilizes the calculation with
respect to renormalization/factorization scales variations, and 
increases slightly the coefficient function in the large-$x$ region.
These findings can be useful for improved phenomenological analyses and
determinations of the quark fragmentation functions at large $x$.

In the heavy-quark case, the cross sections can be computed by convoluting
coefficient functions, evaluated in the massless-quark approximation,
with the (process-independent) perturbative fragmentation function of the 
heavy quark (see Eq.~(\ref{sighq})).
We have shown how the 
initial condition $D^{\rm ini}(x)$, at the scale $m$, for the evolution of 
the perturbative fragmentation function can be obtained from the universal
factorization properties of parton radiation in the collinear and 
quasi-collinear limits. We have then exploited this method to perform 
all-order resummation of the large soft-gluon contributions that
dominate the behaviour of $D^{\rm ini}(x)$ at high~$x$. We have derived
an expression (see Eq.~(\ref{resfin-dini})) for the initial
condition $D^{\rm ini}(x)$ that explicitly resums Sudakov terms
up to NLL accuracy and is consistently matched with the 
complete NLO calculation. 
This result extends 
the LL process-independent calculation of Ref.~\cite{Mele:1991cw} 
to NLL order 
and the NLL process-dependent calculation of Ref.~\cite{Dokshitzer:1996ev}
in a process-independent way.

Our numerical studies of heavy-quark fragmentation show that 
Sudakov resummation softens the high-$x$ behaviour of heavy-quark distributions.
It also leads to a marked
reduction of the dependence on the renormalization/factorization scales,
making the perturbative predictions for heavy-quark fragmentation processes
more reliable. These perturbative features are important for improved 
studies of the non-perturbative component that has to be 
included in phenomenological applications to heavy-quark fragmentation.

\vskip 15pt


\begin{appendix}

\noindent{\bf \Large Appendices}
\section{Two-particle inclusive distribution \\
in $e^+e^-$ annihilation}
\label{sec:2ee}

\begin{figure}[t]
\begin{center}
\begin{picture}(130,100)(0,0)         
\ArrowLine           ( 0,10)( 40,50)         
\ArrowLine           ( 0,90)( 40,50)
\Photon              ( 40,50)( 90,50){3}{4}  
\ArrowLine                (90,50)(120,90)
\ArrowLine                (90,50)(120,10)
\ArrowLine                (90,50)(133,60)
\Line                     (90,50)(135,50)
\ArrowLine                (90,50)(133,40)
\BCirc               ( 90,50){10}
\Text                ( 120,10)[l]{ $h_B(p_2)$}
\Text                ( 125,90)[l]{$h_A(p_1)$}
\Text                ( 135,50)[l]{ $X$}
\Text                ( 45,65)[l]{$V(Q)$}
\Text                ( 5,65)[l]{$e^+$}
\Text                ( 5,35)[l]{$e^-$}
\end{picture}
\caption{\label{epem2fig}\small Inclusive production of two hadrons with 
momenta $p_1$ and $p_2$ in $e^+e^-$ annihilation.
}
\end{center}
\end{figure}
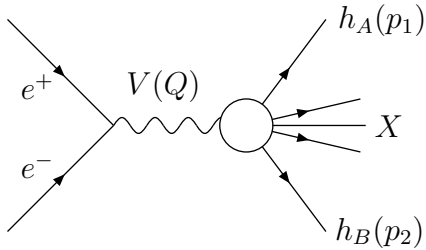

In this appendix we study the less inclusive case of $e^+e^-$ annihilation 
with two observed hadrons, $h_A$ and $h_B$, in the final state
(Fig.~\ref{epem2fig}): 
\beq
\label{gamma12x}
e^+ + e^- \to V(Q)\to h_A(p_1) + h_B(p_2) + X \;\;.
\eeq
We consider the corresponding two-particle inclusive distribution,
\beq
\label{dsigmadx1dx2}
{\cal D}(x_1,x_2;Q^2) \equiv 
\frac{1}{\sigma_{\rm tot}}\frac{d\sigma_{h_Ah_B}^{(e^+e^-)}}{dx_1dx_2} \;\;,
\eeq
where $\sigma_{\rm tot}$ is the total hadronic cross section, and
the kinematic variables $x_1$ and $x_2$ are defined\footnote{Since $x_1$ 
and $x_2$ are not symmetric variables with respect to 
$p_1 \leftrightarrow p_2$, a corresponding 
symmetrization, $h_A \leftrightarrow h_B$, is understood in the definition of 
the two-particle distribution in Eq.~(\ref{dsigmadx1dx2}).}
according to Ref.~\cite{aemp}:
\beq
\label{x1x2}
x_1 \equiv  \frac {2p_1 \cdot Q}{Q^2}\;,\quad x_2 \equiv  
\frac {p_1 \cdot p_2}{p_1 \cdot Q} \;.
\eeq
Note that $x_1$ and $x_2$ can independently vary in the whole kinematical range
between $0$ and $1$. In the \ee centre-of-mass frame, 
$x_1$ coincides with the energy fraction of the hadron with momentum $p_1$, 
while $x_2$ is the energy fraction of the hadron with momentum $p_2$
times $1-\cos \theta_{12}$, $\theta_{12}$ being the angle between the two
hadrons. The choice of these two kinematic variables guarantees
that the two-particle distribution is collinear safe\footnote{If one uses the
energy fractions $2p_1Q/Q^2, 2p_2Q/Q^2$ of the two hadrons in the definition 
of the distribution, an additional physical cutoff has to be introduced
to avoid the kinematic region where $p_1$ and $p_2$ are collinear.} 
with respect to final-state QCD radiation for any values $x_1,x_2 \neq 0$. 
Considering sufficiently large values of both $x_1$ and $x_2$,
we ensure that $\theta_{12}$ is large and, hence, the two observed hadrons 
belong to two different jets in a nearly back-to-back configuration.

More precisely, owing to our definition of the kinematic variables
$x_1$ and $x_2$, the two-particle distribution ${\cal D}(x_1,x_2;Q^2)$
can be computed according to the following QCD factorization 
formula\footnote{The ratio $\sigma^{(LO)}/\sigma_{\rm tot}$ is introduced on 
the right-hand side to make the notation in Eq.~(\ref{ds2had}) consistent with 
that in Eq.~(\ref{dshad}).}
\beeq 
\label{ds2had}
{\cal D}(x_1,x_2;Q^2) &=& \frac{\sigma^{(LO)}}{\sigma_{\rm tot}}
\int_{x_1}^1 \frac{dz_1}{z_1}
\int_{x_2}^1 \frac{dz_2}{z_2} \;
C^{(e^+e^-)}(x_1/z_1,x_2/z_2,\as(\mu^2);Q^2,\mu^2,\mu_F^2) \nonumber \\
&\cdot& D(z_1,\mu_F^2) \;D(z_2,\mu_F^2) \;\;,
\eeeq
where $D(x,\mu_F^2)$ is the same fragmentation function as appears in
Eq.~(\ref{dshad}). The two-particle coefficient function 
$C^{(e^+e^-)}(x_1,x_2)$ is
computable in QCD perturbation theory. In the \naive\ parton model (i.e. at 
the LO), we have $C^{(e^+e^-)}(x_1,x_2)= \delta(1-x_1) \,\delta(1-x_2)$
and, thus, the two-particle distribution is simply given by the product
$D(x_1,Q^2) D(x_2,Q^2) \propto (d\sigma/dx_1) (d\sigma/dx_2)$ of the two 
corresponding single-particle distributions in Eq.~(\ref{dshad}). Higher-order
perturbative contributions lead to QCD correlation effects that spoil this
\naive\ factorized structure and whose size is measured by the 
two-particle coefficient function.

The complete evaluation of $C^{(e^+e^-)}(x_1,x_2)$ at the NLO was performed in
\cite{aemp}. As in the case of the single-particle coefficient function
in Eq.~(\ref{dxqq0}), the flavour non-singlet contributions to 
$C^{(e^+e^-)}(x_1,x_2)$ contain logarithmically-enhanced terms in 
the semi-inclusive limit $x_1,x_2 \to 1$. These terms are conveniently singled
out by introducing double Mellin moments as
\beq
\label{dnn}
{\cal D}_{N_1N_2}(Q^2) \equiv \int_0^1\;dx_1\;x_1^{N_1-1}\int_0^1\;dx_2
\;x_2^{N_2-1}\;
{\cal D}(x_1,x_2;Q^2) \;\;,
\eeq
and studying their large-$N_i$ $(i=1,2)$ behaviour. In $N$-moment space
Eq.~(\ref{ds2had}) becomes
\beeq 
\label{ds2n}
{\cal D}_{N_1N_2}(Q^2) = \frac{\sigma^{(LO)}}{\sigma_{\rm tot}} \;
C_{N_1N_2}^{(e^+e^-)}(\as(\mu^2);Q^2,\mu^2,\mu_F^2) 
\; D_{N_1}(\mu_F^2) \;D_{N_2}(\mu_F^2) \;\;,
\eeeq
and in the limit $N_i \to \infty$ $(i=1,2)$ the coefficient function
$C_{N_1N_2}^{(e^+e^-)}$ has a perturbative expansion similar to 
Eq.~(\ref{gensig}), with $\as^n \ln^mN$ replaced
by $\as^n \ln^{m_1}N_1 \ln^{m_2}N_2$ ($m_1+m_2 \leq 2n$). 

All-order resummation of the logarithmically-enhanced terms can be performed
by using standard techniques. In particular, we can exploit the strict
kinematical analogy with the {\em differential} Drell--Yan (DY) distribution,
and we can proceed as in Sect.~4 of the first paper in Ref.~\cite{ct2}.
We obtain the exponentiated result
\begin{eqnarray}
\label{cn1n2}
&&\!\!\!\!\!\! \!\!\!\!C_{N_1N_2}^{(e^+e^-)}(\as(\mu^2);Q^2,\mu^2,\mu_F^2)
= \exp \left\{ 
\left[ \int_0^1 dx \frac{x^{N_1-1}-1}{1-x}\int_{\mu_F^2}^{(1-x)Q^2}
\frac{dq^2}{q^2}A[\as(q^2)]+\;(N_1 \leftrightarrow N_2) \right] \right.
\nonumber \\
&&\!\!+ \left. \int_0^1\;dx_1 \int_0^1\;dx_2
\frac{(x_1^{N_1-1}-1)}{1-x_1}
\frac{(x_2^{N_2-1}-1)}{1-x_2} \;A[\as((1-x_1)(1-x_2)Q^2)] \right. \\
&&\!\!+ \left. {\cal O}\Bigl(\as(\as \ln N_i)^n\Bigr) 
\right\} \;\;. \nonumber
\end{eqnarray}
The structure of Eq.~(\ref{cn1n2}) is similar to that of 
Eqs.~(\ref{cnee})--(\ref{jfun})
for the single-particle coefficient function. 
The main difference is 
that the integrands of Eq.~(\ref{cn1n2}) contain a single perturbative
function, namely the soft-gluon function $A(\as)$ given by Eq.~(\ref{afun}). 
This is due to the fact that, when $x_1$
and $x_2$ are both large, only the emission of soft gluons 
is kinematically allowed in the final state.

The perturbative expansion of Eq.~(\ref{cn1n2}) at ${\cal O}(\as)$
agrees with the NLO result of Ref.~\cite{aemp}.
Owing to the knowledge of the first two coefficients (see Eq.~(\ref{a1a2}))
of the function $A(\as)$, the result in Eq.~(\ref{cn1n2}) resums the
LL ($m_1+m_2 = n+1$) and NLL ($m_1+m_2 = n$)
terms $\as^n \ln^{m_1}N_1 \ln^{m_2}N_2$ in the exponent of the
two-particle coefficient function $C_{N_1N_2}^{(e^+e^-)}$. 
The integrals in Eq.~(\ref{cn1n2}) can explicitly be performed up to 
NLL accuracy, and the resummed coefficient function can be written
in the following equivalent form:
\beeq
C_{N_1N_2}^{(e^+e^-)}(\as(\mu^2);Q^2,\mu^2,\mu_F^2)
&=& \exp \Bigl[ \; \left( \ln N_1 + \ln N_2 \right) 
\;g_{2p}^{(1)}(\lambda_{12}) 
\nonumber \\
&+& g_{2p}^{(2)}(\lambda_{12},Q^2/\mu^2;Q^2/\mu_F^2) 
+ {\cal O}(\as(\as \ln N_i)^n) \Bigr] \;,
\eeeq
where
\beq
\lambda_{12} = b_0 \;\as(\mu^2) \left( \ln N_1 + \ln N_2 \right) \;,
\eeq
and the LL and NLL functions $g_{2p}^{(1)}$ and $g_{2p}^{(2)}$ are given by 
\beeq
\label{g1fun2p} 
g_{2p}^{(1)}(\lambda_{12}) &=& 
\frac{A^{(1)}}{\pi b_0 \lambda_{12}} \;
\bigl[ \lambda_{12} + (1-\lambda_{12}) \ln (1-\lambda_{12}) \bigr] \;,\\
g_{2p}^{(2)}(\lambda_{12},Q^2/\mu^2;Q^2/\mu_F^2) 
&=& \frac{A^{(1)}  b_1}{2 \pi b_0^3}
\left[ 2\lambda_{12} + 2 \ln (1-\lambda_{12}) +  \ln^2 (1-\lambda_{12}) \right]
\nn \\
\label{g2fun2p}
&-&\frac{1}{\pi b_0} \left[\lambda_{12} + \ln (1-\lambda_{12}) \right] 
\left( \frac{A^{(2)}}{\pi b_0} - A^{(1)} \ln \frac{Q^2}{\mu^2} \right)  \nn \\
&-& \frac{2 A^{(1)}\gamma_E}{\pi b_0} \ln (1-\lambda_{12})
- \frac{A^{(1)}}{\pi b_0} \;\lambda_{12} \;\ln \frac{Q^2}{\mu_F^2} \;. 
\eeeq

The coefficient functions on the right-hand side of the factorization formulae 
in Eqs.~(\ref{dshad}) and (\ref{ds2had})
are factorization-scheme-dependent. The resummed expressions in 
Eqs.~(\ref{eeall}) and (\ref{cn1n2}) are valid in the 
\ms\ factorization scheme. The ratio 
\begin{equation}
\label{deln1n2}
\Delta^{(e^+e^-)}_{N_1 N_2}(\as(\mu^2);Q^2,\mu^2)\equiv 
\frac{C_{N_1N_2}^{(e^+e^-)}(\as(\mu^2);Q^2,\mu^2,\mu_F^2)}
{C_{N_1}^{(e^+e^-)}(\as(\mu^2);Q^2,\mu^2,\mu_F^2) 
\;C_{N_2}^{(e^+e^-)}(\as(\mu^2);Q^2,\mu^2,\mu_F^2)}
\end{equation}
between two-particle and one-particle coefficient functions
is instead independent of the factorization scheme. This ratio corresponds
to the two-particle coefficient function as defined in the alternative
factorization scheme\footnote{The annihilation scheme amounts to redefining
the \ms\ fragmentation function in such a way that the corresponding
coefficient function in Eq.~(\ref{dshad}) is equal to unity.}, 
named `annihilation scheme'~\cite{Nason:1994xx}, 
introduced in Ref.~\cite{aemp}. Using our NLL resummed expressions for
$C_{N_1N_2}^{(e^+e^-)}$ and $C_{N}^{(e^+e^-)}$, it is straightforward to check
that the dependence on the factorization scale $\mu_F$ consistently cancels
in the ratio $\Delta^{(e^+e^-)}_{N_1 N_2}$, i.e. in the right-hand side
of Eq.~(\ref{deln1n2}).

\section{Single-particle inclusive distribution \\
in lepton--hadron collisions
}
\vskip 10.pt

\begin{figure}[t]
\begin{center}
\begin{picture}(130,100)(0,0)  
\ArrowLine           ( 0,100)( 60,80)
\ArrowLine           ( 60,80)( 120,100)
\Photon              ( 60,80)( 60,40){3}{4}
\ArrowLine           ( 0,10)( 60,40) 
\ArrowLine           ( 60,40)(120,40)   
\Line           ( 60,40)(110,10)
\ArrowLine      ( 60,40)(100,10) 
\ArrowLine      ( 60,40)(110,15)  
\BCirc               ( 60,40){10}
\Text                ( 115,15)[l]{ $X$}
\Text                ( 110,110)[l]{$l'$}
\Text                ( 125,45)[l]{ $h'(k)$}
\Text                ( 25,65)[l]{$V(q)$}
\Text                ( 5,110)[l]{$l$}
\Text                ( 5,30)[l]{$h(p)$}
\end{picture}
\caption{\label{disfig}\small Inclusive production of a single hadron $h'$ with
momentum $k$ in deep inelastic lepton--hadron scattering.
}
\end{center}
\end{figure}
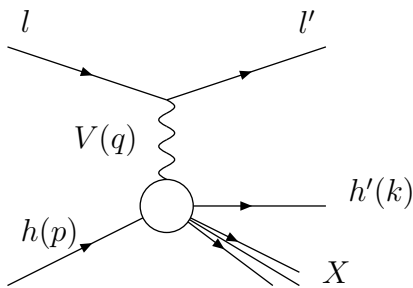

We conclude our analysis of light-hadron fragmentation in the semi-inclusive
region by considering deep inelastic lepton--hadron scattering (DIS).
Instead of studying the DIS total cross section, we are interested in the 
inclusive production of a single final-state hadron (Fig.~\ref{disfig}):
\begin{eqnarray}\label{gpkx}
l + h(p) \rightarrow l' +h'(k) + X \;\;.
\end{eqnarray}
To study the fragmentation of the tagged hadron at high momentum
fraction, we can limit ourselves to considering the approximation in which the
scattering process in Eq.~(\ref{gpkx}) occurs through the exchange of a single
vector boson with momentum $q$ and hardness $-q^2=Q^2 > 0$. We denote by
$p$ and $k$ the momenta of the incoming and final-state hadrons, respectively.

Besides the customary Bjorken variable $x_B$, we define the final-state
variable $z$ as follows~\cite{aemp}:
\begin{eqnarray}\label{xbz}
x_B \equiv \frac{Q^2}{2p \cdot q} \; , \quad z \equiv \frac{p \cdot k}
{p\cdot q} \; ,
\end{eqnarray}
and the semi-inclusive region we are interested in is 
specified by the limit $x_B \to 1$ and $z \to 1$.

In the Breit frame, where $q^{\mu}=(0,{\bf 0}, -Q)$ and 
$p^{\mu}=(1,{\bf 0},1) Q/(2x_B)$, we have $z=(1-\cos\theta)k_0/Q$, so 
the variable $z$ is related to the energy
$k_0$ of the fragmenting hadron and to its scattering angle $\theta$
with respect to $p$.  
When approaching the semi-inclusive limit, the two hadrons $h(p)$ and $h'(k)$
are in a back-to-back configuration, with the target hadron momentum
$p^{\mu}\simeq (1,{\bf 0},1) Q/2$ going forward and the fragmenting hadron
momentum $k^{\mu}\simeq (1,{\bf 0},-1) Q/2$ moving backward in the current-jet
hemisphere.

Note that $x_B$ and $z$ can independently vary in the whole kinematical range
between $0$ and $1$. Moreover, as long as $z$ is not vanishing, the fragmenting
hadron $h'(k)$ cannot become collinear to the colliding hadron $h(p)$.
Therefore, the single-inclusive cross section 
$d\sigma_{h'}^{(\rm DIS)}/dx_B dz$ fulfils \cite{aemp} the following
QCD factorization formula:
\beeq 
\label{dsigdxdz}
\frac{d\sigma_{h'}^{(\rm DIS)}}{dx_B\;dz}
&=& \sigma^{(0)} 
\int_{x_B}^1 \frac{dx}{x} \int_{z}^1 \frac{dy}{y} \;
C^{(\rm DIS)}(x,y,\as(\mu^2);Q^2,\mu^2,\mu_F^2) \nonumber \\
&\cdot& F(x_B/x,\mu_F^2) \;D(z/y,\mu_F^2) \;\;,
\eeeq
where $\sigma^{(0)}$ is the LO cross section, $D(z,\mu_F^2)$ is the parton
fragmentation function of the tagged hadron $h'(k)$ and $F(x,\mu_F^2)$
is the parton distribution function of the colliding hadron $h(p)$.
To be defined, we recall \cite{book} that $F(x,\mu_F^2)$ appears in the
analogous factorization formula for the DIS total cross section
$d\sigma^{(\rm DIS)}/dx_B$:
\beq 
\label{dsigdx}
\frac{d\sigma^{(\rm DIS)}}{dx_B} = \sigma^{(0)} 
\int_{x_B}^1 \frac{dx}{x} \;C^{(\rm DIS)}(x,\as(\mu^2);Q^2,\mu^2,\mu_F^2) 
\;  F(x_B/x,\mu_F^2) \;\;.
\eeq
Since eventually we are mainly interested in the semi-inclusive limit, 
we omit parton indices in Eqs.~(\ref{dsigdxdz}) and (\ref{dsigdx}), 
and we understand that the parton fragmentation function and distribution 
function refer to the flavour non-singlet components.

The single-inclusive coefficient function $C^{(\rm DIS)}(x,y)$ is 
computable in QCD perturbation theory.
In the \naive\ parton model (i.e. at 
the LO), we have $C^{(\rm DIS)}(x,y)= \delta(1-x) \,\delta(1-y)$,
and the single-inclusive cross section in Eq.~(\ref{dsigdxdz}) is simply 
proportional to the product of the parton distribution and the fragmentation 
function,
$d\sigma_{h'}^{(\rm DIS)}/dx_B dz \propto F(x_B,Q^2) D(z,Q^2)$.
The complete NLO calculation of $C^{(\rm DIS)}(x,z)$ was performed in 
Ref.~\cite{aemp}. 

As in the case of the $e^+e^-$ two-particle coefficient function
$C^{(e^+e^-)}(x_1,x_2)$ in Eq.~(\ref{ds2had}), at high perturbative orders
the flavour non-singlet contribution to $C^{(\rm DIS)}(x_B,z)$ 
contains terms that are logarithmically enhanced in 
the semi-inclusive limit $x_B,z \to 1$. Similarly to the two-particle 
distribution in $e^+e^-$ annihilation, it is thus convenient to define the 
double Mellin moments: 
\begin{eqnarray}\label{ffnn}
C^{(\rm DIS)}_{N_1N_2}(\as(\mu^2);Q^2,\mu^2,\mu_F^2) \equiv \int_0^1dx\; x^{N_1-1}\int_0^1dz\; 
z^{N_2-1} C^{(\rm DIS)}(x,z,\as(\mu^2);Q^2,\mu^2,\mu_F^2) \; ,
\end{eqnarray}
and to consider their large-$N_i$ $(i=1,2)$ behaviour.

The general structure of $C^{(\rm DIS)}_{N_1N_2}$ in the large-$N_i$ limit is 
similar
to that of $C^{(e^+e^-)}_{N_1N_2}$ and the resummation of the large logarithmic 
contributions can be performed following Ref.~\cite{ct2}. 
Using the \ms\ factorization scheme, to NLL order we find
\begin{eqnarray}\label{lnffnn}
\ln C^{(\rm DIS)}_{N_1N_2}(\as(\mu^2);Q^2,\mu^2,\mu_F^2)=
\ln C^{(e^+e^-)}_{N_1N_2}(\as(\mu^2);Q^2,\mu^2,\mu_F^2)
+ {\cal O}(\as(\as\ln N_i)^n) \;\;,
\end{eqnarray}
with $C^{(e^+e^-)}_{N_1N_2}$ given by the expression in Eq.~(\ref{cn1n2}).
This result has a simple physical explanation.
In the semi-inclusive limit, the kinematical configuration of the DIS
single-inclusive cross section is related to that of the $e^+e^-$
two-particle cross section by crossing the DIS incoming hadron to the
final state. In both cases the cross section is dominated by purely soft
emission. The relation (\ref{lnffnn}) thus follows from the fact that the
intensities of soft-gluon radiation from space-like and time-like partons are
equal up to NLL accuracy 
(see Ref.~\cite{cmw} and the discussion at the end of Sect.~\ref{1ee}).

\end{appendix}

\vskip 10pt

\noindent {\bf Acknowledgments.} We wish to thank Luca Trentadue for
participation in early stages of this work. M.C. thanks Ugo
Aglietti for many conversations on this subject.

\end{document}